\documentclass{elsart}

 \usepackage{graphicx}

\usepackage{amssymb}

\usepackage{latexsym}
\usepackage{ifthen}

\newcommand\eps{\epsilon}

\newcommand\obs{A}

\newcommand\be{{\bf e}}
\newcommand\bx{{\bf x}}
\newcommand\bu{{\bf u}}

\newcommand\bs{{\bf s}}

\newcommand\bxi{\mbox{{\boldmath $\xi$}}}

\newcommand\bF{{\bf F}}

\newcommand\bU{{\bf U}}
\newcommand\bS{{\bf S}}
\newcommand\bX{{\bf X}}

\newcommand{\rfq}[1]{~(\ref{eq:#1})}
\newcommand{\mod}{\mbox{\rm mod}\: }
\renewcommand{\div}{\mbox{\rm div} }

\begin{document}

\begin{frontmatter}



\title{Linear response,  susceptibility and resonances  in  chaotic toy models}


\author{Bruno Cessac and Jacques-Alexandre Sepulchre}

\address{Institut Nonlin\'eaire de Nice \\ 1361 route des Lucioles,  06560 Valbonne, France}
  
\begin{abstract}
We consider simple examples  illustrating some new features of  the linear response theory developed by Ruelle for
 dissipative  and chaotic systems~[{\em J. of Stat. Phys.}  {\bf 95}  (1999)  393].   In this theory the  concepts of linear response,
 susceptibility and resonance, which are familiar to physicists,  have been  revisited due to the dynamical  contraction of the whole 
 phase space onto attractors.  In particular the standard framework of the ``fluctuation-dissipation" theorem  breaks down and new 
 resonances  can show up oustside the powerspectrum.   In previous papers we  proposed and used new  numerical methods to demonstrate the
 presence of the new resonances predicted by Ruelle in a model of chaotic neural network.   In this article  we deal with  simpler models
 which can be   worked out  analytically in  order to gain more insights into  the genesis of the ``stable'' resonances and their
 consequences on the  linear response of the system.  We consider a class of 2-dimensional time-discrete maps   describing simple rotator
 models  with  a contracting radial dynamics onto  the unit circle and a  chaotic angular dynamics  $\theta_{t+1} = 2 \theta_t (\mod 2\pi)$.
  A generalisation of this system to a network of interconnected rotators is also analysed
 and related with our previous studies \cite{CS1,CS2}.
These  models   permit us to classify the different  types of resonances in the susceptibility and  to discuss in particular the relation between  the relaxation time of the system to equilibrium   with the {\em mixing}  time given by the decay of the correlation functions.   Also it enables one
to  propose some general mechanisms responsible for the creation of stable resonances with arbitrary frequencies, widths, and  dependency on the pair of perturbed/observed variables.   
\end{abstract}

\end{frontmatter}



\section{Introduction}

In statistical physics, a standard idea is that the linear response of a system can be computed in terms of correlation functions of this system at equilibrium. This is the cornerstone of important results like the fluctuation-dissipation theorem~\cite{Landau,Dorfmann}.   More recently a general theory for the linear response  for hyperbolic  systems has been developed by Ruelle~\cite{Ru98,Ru99}.   This theory shows that   the  picture of correlation functions  is  incomplete in this case.   In this situation a linear response function has in general two contributions, corresponding  respectively to the expanding (unstable)  and the contracting (stable) 
directions  of the hyperbolic dynamics.   Although the first contribution, called in the sequel the {\em unstable}  
response function,  can still  be associated 
with some correlation function,  this is not true  for the second contribution, named the {\em stable} response 
function\footnote{Though the terminology ``stable'', ``unstable'' is a little bit
confusing when dealing with linear response and resonances, we use it because
it refers explicitely to unstable and stable spaces in the tangent space of the
attractor. The mechanism leading to the existence of a stable and unstable contribution of the linear response
are of completely different nature: this is exponential contraction
for the stable part and this is exponential mixing for the
unstable part.}.   This property   transfers  to  the complex susceptibility, i.e. the frequency-dependent responses of the system, for which we can  also  distinguish  between two types of  resonances,
namely the  stable and the unstable resonances.

Computations  of the linear response in specific  chaotic systems have been performed by several authors~\cite{FIV}-\cite{Reick}.   For example  in~\cite{FIV},  Falcioni {\it et al.}  discuss and numerically apply  a simple method to calculate  the (impulse) linear  response function.   The procedure is an averaging  on perturbations of the system  by means of  Dirac kicks.
 However, as estimated by the authors, the convergence of this method  is slow  in the time-domain.   Another study has been performed  by Reick in~\cite{Reick},   who uses a different technique,  averaging the effects of  time-periodic perturbations.  With this method he numerically demonstrates  the existence of  linear response in  the Lorentz system,  although the latter is known to be  nonhyperbolic.   Neither of these previous works, however,  obtain evidence of the stable resonances predicted by Ruelle.    Let us remark  that contrary to the present study, these papers do not  consider chaotic systems with state-dependent  perturbations. 

 In a recent work  we have illustrated for the first time the  prediction of Ruelle on a simple model of networks of interconnected units exhibiting  chaotic dynamics~\cite{CS1}.    Using  independently the same averaging procedure as~\cite{Reick},  we managed  to  compute the  susceptibility $\hat{\chi}_{ij}(\omega)$ and its complex poles.  (Here $\hat{\chi}_{ij}(\omega)$ denotes the complex amplitude of the average response of unit $i$ to a harmonic excitation of unit $j$ at frequency $\omega$).     
Our numerical method could not  provide separately   the unstable and the stable contributions in the susceptibility, but our  results demonstrated   the existence of stable poles in $\hat{\chi}_{ij}(\omega)$  by  setting in evidence   poles which were  not  in the set of  poles of the correlation functions (the Ruelle-Pollicott resonances~\cite{Gaspard,Policott}).  

On the other hand  our  numerical results pointed out towards  interesting  features whose consequences are worth to explore further.   First,   stable resonances in $\hat{\chi}_{ij}$ can be specific to the pair $ij$.   This property, which is not true in general for the Ruelle-Pollicott resonances,  is quite useful for networks  as it can be used to transmit an  amplitude modulated signal between two specific units in the network~\cite{CS2}.  Another  unexpected   feature is that the average relaxation time of unit $i$ to an impulse perturbation of unit $j$  (which can be predicted  by  Fourier transform of $\hat{\chi}_{ij}(\omega)$)   can extend on  a much longer time lapse  than the {\em mixing}  time, i.e. the decay time  associated with the correlation functions. 
 
  These properties cannot be understood   in the framework of the classical fluctuation-response theorem.  Indeed they reflect the presence of the  stable contribution of the response functions which are not correlation functions.  However the stable response or the stable susceptibility  are  difficult to extract  numerically.   Therefore the  goal of the present  paper is to study  a class of toy models for which  the stable and unstable decomposition of response functions can be worked out  explicitly.   In particular these examples clearly show the relative contributions of correlation and non-correlation terms in the response functions,  and their treatment helps  to better understand the above  properties revealed  in networks.
   These simple examples  are also treated numerically.
  This allows one to validate the numerical procedure used in~\cite{CS1,CS2}. 
  
In the next Section we  recall some elements of Ruelle's theory which will be extensively used later on. We formally derive  his general response formula by using the method of the  impulse response and then introduce the definition of susceptibility as being its Fourier transform.  In Section~3 we treat  the basic example $\theta_{t+1} = 2 \theta_t$ (mod $2 \pi$)  in this framework  and discuss briefly  the issue of the non-invertibility of this dynamics.
 This elementary example   serves also to describe and to test numerical methods.     Section~4 concerns  class of uniformly hyperbolic systems  which modelise  chaotic   rotators in the plane.   Various extensions of the simplest model are considered in the following subsections so as to discuss properties of the stable resonances. They are summarised in the Conclusions.  
 An appendix of the paper  presents a natural extension  of our chaotic rotator model in higher dimensions for which some properties generalise directly.

\section{A general linear response formula of Ruelle}
\label{sec:general}

We start by reviewing  some results of the linear response theory obtained  by Ruelle~\cite{Ru98,Ru99}. We concentrate on  the particular case of autonomous dynamical systems described by iterations of a map.  Consider the following systems whose dynamics is governed by the recurrence equation:
\begin{equation}
\label{eq:F}
\bx_{t+1}=\bF(\bx_t)
\end{equation} 
where $t\in \mathbb{Z},  \:   \bx \in M $, and $M$ is the phase space, e.g.  a compact manifold in $ \mathbb{R}^N $.   The map is defined by a smooth function  $ \bF(\bx) $,  not necessarily invertible,  and $\bx_t$ denotes the state of system at time $t$.  We also use  the notation  $\bF(\bx_t) = \bF^t(\bx_0)$.
The dynamics of~\rfq{F} is assumed to be chaotic, mixing  and associated with an ergodic measure  $ \rho_F $ of  Sinai-Ruelle-Bowen type (SRB)~\cite{Ru99,topol}.
This implies that any measurable  observable $\obs(\bx) $  has a  time average
\begin{equation}
\overline{A} = \lim_{T \rightarrow \infty} \frac{1}{T} \sum_{t=0}^T \, A( \bF^t(\bx))
\label{eq:Abar}
\end{equation}
which is equal to  its  ensemble average:
\begin{equation}
\label{eq:meanB}
 < \obs > = \int \rho _F (d\bx) \obs(\bx) = \lim_{t \rightarrow \infty}  \int \! d\bx \, \obs(\bF^t(\bx))
\end{equation} 
for almost every initial conditions $\bx$ selected with respect to the Lebesgue measure $d\bx$.  Note that in  eq.\rfq{meanB},  $\rho_F$  equivalently  denotes  the image of the Lebesgue measure under $F^t$, i.e.
$\rho_F = \lim_{t \rightarrow \infty}  F^{*t} \; d\bx$ where the system is assumed to be mixing. 

\subsubsection*{Example:}  We consider a first model  which will be dealt with in next Sections. This  simplest example for system\rfq{F} is given by the mapping:
\begin{equation}
\theta_{t+1} =  f(\theta_t)
\label{eq:f}
\end{equation}
 with $f: \theta \mapsto  2 \, \theta \: (\mod \; 2\pi)$  defined on the unit circle.  This system is chaotic with Lyapunov exponent $\log 2$.  This means that a small perturbation $\delta\theta$ of  any initial condition  $\theta$ is locally amplified in time with speed $2^t$.  On the other hand, {\em on} the unit circle, the difference between the initial trajectory and the perturbed one  is blurred in time. In other words, despite the sensitivity to initial condition, the {\em mean} effect of the perturbation $\delta\theta$ applied at time $t=0$ only  on observables of this system vanishes in time.  This is readily seen as a consequence of  the ergodic property $\overline{A} = <A>$, namely:
\[
 \lim_{T \rightarrow \infty} \frac{1}{T} \sum_{t=0}^T \, A( f^t(\theta+\delta\theta)) =  <A>
\]
for almost every initial conditions.   Let us recall that in this example the SRB measure is $d\theta/2\pi$ and so $<A> = \frac{1}{2\pi} \int_0^{2\pi} \! d\theta \, A(\theta)$. 

\smallskip

Therefore in the class of  chaotic systems considered here, a single perturbation of  initial condition has no effect on the mean value of any observable, although the short time effect of this perturbation can be important.  
Now let us suppose that one applies on system\rfq{F}   a permanent  perturbation depending  on  time $t$ and possibly  on the state $\bx_t$ of this system.  Then the goal of the response theory is to study  how the mean values of observables are affected by this change, compared with the unperturbed system.   This problem  is not easy  in general  because in this case the SRB state is asymptotically different from the one of the unperturbed system.

Thus now let us consider a perturbed version of system\rfq{F}  governed by the following equation:
 \begin{equation}
\label{eq:deltaF}
\tilde{\bx}_{t+1}=  \tilde{\bF}_t(\tilde{\bx}_t)
 \stackrel{\mbox{\rm def}}{=}      \bF(\tilde{\bx}_t) + \bxi_{t+1} (\tilde{\bx_t})
 \end{equation}
The term  $\bxi_{t+1}(\bx)$  defines  the perturbation\footnote{The subscript $t+1$ in eq.\rfq{deltaF} is neither a missprint neither a violation of causality.  The motivation to denote the  perturbation acting on $x_t$  by $\bxi_{t+1} $ instead of $\bxi_{t}$ is that it allows one to write  the subsequent convolution formula\rfq{Ruelle}  and the Fourier transform\rfq{FT}  into the standard  forms they have for continuous time systems}.
The iterate at time $t$ of initial condition $\bx_s$  taken at time $s$ will be denoted
$ \tilde{\bx}_t = \tilde{\bF}^{(t,s)}(\tilde{\bx}_s)$.  (In the following we will abandon  the tilde notation on $\bx$ when no confusion is possible). 
  Then one can  define the mean value of observable $A$ with respect to the perturbed system as~\cite{Ru98}: 
\begin{equation}
< A >_t  = \lim_{s \rightarrow -\infty}  \int  \, d\bx \, A(\tilde{\bF}^{(t,s)}(\bx))
\label{eq:meanAt}
\end{equation}

\subsection{The impulse linear response}

 The goal of the  {\em linear} response theory aims to compute   $  <\obs>_t  - <\obs>$   to first order in $\bxi$.  This will be denoted by $< \delta \obs>_t $.  The relevance of this concept  in chaotic  dynamical systems has been analysed  by Ruelle.  The latter introduced  the concept of differentiation of a  SRB state in~\cite{Ru97} and provided general formula for its computation in~\cite{Ru98,Ru99}
under the asumption of uniform hyperbolicity.
 
 We   (formally)   derive the formula  of Ruelle in a different way by using the method of  the  impulse perturbation.  Consider  a perturbation of the form $ \bxi_t (\bx) = \bxi(\bx) \, \delta_{t\tau}  $.   In this case,  and if  $t \geq \tau > s$ one can write:
\[
\tilde{\bF}^{(t,s)} =  \tilde{\bF}_{t-1}\circ \cdots \circ \tilde{\bF}_{\tau-1} \cdots  \circ \tilde{\bF}_{s} = \bF^{t-\tau}\circ (\bF+\bxi) \circ \bF^{\tau-s-1}
\]
since $\tilde{\bF}_{t} = \bF$ for  all $t$ except for $\tilde{\bF}_{\tau-1} = \bF +  \bxi$. 
Therefore eq.\rfq{meanAt} becomes: 
\begin{eqnarray*}
< A >_t & = &  \lim_{s \rightarrow - \infty}  \int  \, d\bx \, \left[ A\circ \bF^{t-\tau}\circ (\bF+\bxi) \circ \bF^{\tau-s-1}\right](\bx)\\
& = &  \int  \, \rho_F(d\bx) \, \left[ A\circ \bF^{t-\tau}\right] (\bF(\bx)+\bxi(\bx))
\end{eqnarray*}
where we have used eq.\rfq{meanB} to take the limit $s \rightarrow -\infty$. 
On the other hand,  by using the invariance of $\rho_F$ under $\bF$, one can write $<A> =  \int  \, \rho_F(d\bx) \, \left[ A\circ \bF^{t-\tau}\right] (\bF(\bx))$  .  Consequently, the  impulse response at time $t> \tau$ can be set under the following form:
 \begin{equation}
   <\obs>_t  - <\obs>  =   \int\rho_F(d\bx) \left[ (\obs \circ \bF^{t-\tau} ) ( \bF(\bx) + \bxi (\bx) ) -  (\obs \circ \bF^{t-\tau} ) (\bF(\bx)) \right]   \label{eq:prop2}
\end{equation}
  Although the expression\rfq{prop2} is valid beyond the linear regime, this form is useful  for computing the linear response $<\delta \obs>_t $.  The latter will be denoted $\chi_{A\xi}(t-\tau)$ and can be readily deduced as the first order Taylor expansion of\rfq{prop2}:
\begin{equation}
\chi_{A\xi}(t-\tau)=  \int\rho_F(d\bx)    \nabla  (\obs \circ \bF^{t-\tau} ) (\bF(\bx)) \cdot  \bxi  (\bx)
\label{eq:ImpulLinResp}
\end{equation}
Now  a general time-dependent perturbation $\bxi_t(\bx)$ can be written  as the sum of impulses
$\bxi_t(\bx) = \sum_{\tau = -\infty}^{t} \bxi_{\tau} (\bx) \, \delta_{t\tau} $. 
 Thus, the linear response is given by a superposition  of terms like\rfq{ImpulLinResp}, namely:
\begin{equation}
\label{eq:Ruelle}
 <\delta \obs>_t  = \sum_{\tau = -\infty}^{t}  \chi_{A\xi}(t-\tau)   \cdot \bxi_{\tau}(\bx) 
\end{equation}
This  general convolution formula is the linear response derived by Ruelle in~\cite{Ru98,Ru99}.
Note that it is not evident that the series converge.  This point was initially  raised in  the famous objection of van Kampen~\cite{vK71}; 
Due to the existence of unstable
directions and positive Lyapunov exponents, one would actually expect that the linear response diverges.  
However,  this issue  was  highlighted  by the  statistical physics arguments of Kubo~\cite{Kubo86} and by several subsequent studies, as the ones  of Falcioni et al.~\cite{FIV} and  the recent contribution of  Bofetta et al.~\cite{BKMV}.   In the present context where one assumes the uniform hyperbolicity, one can follow the argument of Ruelle who has shown, by using projections on the local stable and unstable subspace,  that the unstable part is a correlation function, thus the series converges due to (exponential)
mixing occuring in uniformly hyperbolic systems;
 on the other hand the stable part converges due to exponential contraction.

\subsection{The susceptibility}

From eqs.\rfq{ImpulLinResp}-(\ref{eq:Ruelle}), considering  perturbations of the form $\bxi_t (\bx) = \bxi(\bx) \, e^{-i\omega t}$, it is easy to show that the linear response of (the mean value of) observable $A$ is given by
\[
<\delta \obs>_t  =  \hat{\chi}_{A\xi}(\omega)  \, e^{-i\omega t}
\]
where the complex amplitude
\begin{equation}
\label{eq:FT}
\hat{\chi}_{A\xi}(\omega) = \sum_{t=-\infty}^{\infty} \chi_{A\xi} (t) \, e^{i\omega t}
\end{equation}
is called the {\em susceptibility}.   The latter is nothing but the Fourier transform of the impulse response $\chi_{A\xi} (t)$  of observable $A$ with respect to perturbation  $\bxi$  (with $\chi_{A\xi}(t)=0$  for $t<0$ ).  
Conversely, the impulse response can be computed from the susceptibilty by   inverse Fourier transform  $\chi_{A\xi}(t) = \frac{1}{2\pi} \int_0^{2\pi}  \hat{\chi}_{A\xi}(\omega)  \, e^{- i\omega t} d\omega$.

Our derivation of eq.\rfq{Ruelle} gives  a straightforward but non rigorous way to obtain  the linear response,  and in particular the susceptibility, for dynamical systems of the form\rfq{F}. The rigorous analysis has been done by Ruelle. 
In fact, in refs.~\cite{Ru98,Ru99} using the hypothesis of
uniform  
  hyperbolicity of the systems. Ruelle goes much beyond the  eqs.\rfq{ImpulLinResp}-(\ref{eq:FT})
by proving a general decomposition of the response functions  into two terms --stable and unstable-- which stems from the natural foliation of the phase space into stable and unstable manifolds and its implications for the SRB measure.
 In the following Sections,   instead of rephrasing  Ruelle's theory about  this decomposition proved for a general system,  the goal is to apply this concept  on  simple  systems  in order to figure out  some of its consequences  already observed in our previous works~\cite{CS1,CS2}.  Prior to analyse these effects, we review  in next  paragraph  the simplified  case where the system has only expanding directions and so no contractions in phase space. 
     
 

\section{Purely expanding dynamics.}

\subsection{A basic example.}

We start by considering  the model  of example 1   introduced above ($f: \theta \mapsto  2 \, \theta \: \mod \; 2\pi$),   perturbed by a small periodic forcing $\eps \xi_t  =\eps \xi \, e^{-i\omega t}$, namely:
\begin{equation}
\theta_{t+1} =  f(\theta_t) + \eps\, \xi(\theta_t) e^{-i\omega (t+1)}
\label{eq:fperturb}
\end{equation}
Let us consider the observable $A(\theta) = $ ``the value of  $\theta$   in $[0,2\pi)$''\footnote{Let us remark that whereas  the observable $A(\theta)$  is a natural choice to measure the angle, it  is discontinuous for each $\theta = n 2\pi$  ($n$ integer), with a discontinuity equal to  $-2\pi$ .  Thus the derivative of $A$  standing  in eq.\rfq{ImpulLinResp} should be treated as
$ \partial_{\theta} A =  1 - 2\pi  \sum_n \delta(\theta-n2\pi)$. 
\label{eq:distribution}
 In fact, as it is shown  below, the method of integration by part  avoid the explicit use of this  distribution.}.
   When $\eps = 0$, we have $<A(\theta)> = \pi$ .  
Then, following the discussion of the  previous Section,  the mean value of $A(\theta)$ in the perturbed system  is given by $<A(\theta)>_t  = \pi + \eps \, \hat{\chi}(\omega) e^{-i\omega t}$.   In this expression  the susceptibility $\hat{\chi}(\omega)$ is  the Fourier transform of the impulse response  $\chi(t)$  and the latter can be computed  according to eq.\rfq{ImpulLinResp}  as:
\begin{equation}
\chi (t) = \frac{1}{2\pi}\int_0^{2\pi} \! d\theta \,  \partial_{\theta}(A \circ f^t) (f(\theta)) \, \xi(\theta)
\label{eq:RepContMeas}
\end{equation}
for $t \geq 0$.   The notation $\partial_{\theta} (A \circ  f^t)  (f(\theta))$ means the derivative w.r.t. $\theta$  of $A(f^t(\theta))$  evaluated at $f(\theta)$.
To treat eq.\rfq{RepContMeas} one wishes to use a change of variable $\phi  = f(\theta)$ followed by an integration by part.  This would lead us to the fluctuation response theorem.    Since $f$ is not assumed invertible, the proposed change of variable is not straightforward. We shall come back to the general case later on. For simplicity  we first consider a particular  situation, where the perturbation function $\xi(\theta)$ can be written as $(X\circ f)(\theta)$.
In this case,  a change of variable $\phi  = f(\theta)$ can be performed in\rfq{RepContMeas}.
This change of variable can be followed by an
integration by part  leading  to the following expression (we return to the variable $\theta$):
\begin{equation}
\chi (t)  = X(2\pi) \delta_{t,0} -  \frac{1}{2\pi} \int_0^{2\pi} \! d\theta  \,  A(f^t (\theta))\, \partial_{\theta}X(\theta)
\label{eq:RepCorrel0}
\end{equation}
The first term in the right-hand side of this equation  
 results  from the integrated term $ \frac{1}{2\pi} [A(f^t(\theta)) X(\theta)]^{2\pi}_0$.  This is consistent with the trivial case where  $X$ is a constant.  Then the impulse response is zero except at $t=0$ where it is equal to this constant.  Conversely  the integrated term  disappears for an observable $A$  and a perturbation $X$  continuous  in $\theta$.  In this case one obtains a clearcut application of the fluctuation-response theorem:   the linear response is indeed   a correlation function\footnote{If $\obs(\bx)$ and $B(\bx)$ are two observables, the {\em  time correlation function} of these observables is defined by:
\begin{equation}
\label{eq:correlation}
 < \obs(\bx_t ) ; B(\bx_0))> = \int \rho _F (d\bx) A( \bF^t (\bx))  B (\bx) - <A><B>
\end{equation}
 } (truncated to $0$ for $t<0$). This  can be expressed as:
\begin{equation}
\chi (t)  =  - < \theta_t \, ; \partial_{\theta}X (\theta_0) >
\label{eq:RepCorrel1}
\end{equation}
for $t \geq 0$. 
As a concrete example , let us consider the function 
\begin{equation}
X(\theta) = \frac{6}{(2\pi)^2} \theta(2\pi - \theta)
\label{eq:Xtheta}
\end{equation}
(The number $(6/2\pi)^2$ is just a normalisation factor such that $<X> = 1$.)   Then a simple calculation yields the following function:
\begin{equation}
\chi (t)  = 12  < \theta_t \,; \theta_0 > =    2^{-t}  
\label{eq:RepCo}
\end{equation}
for $t \geq 0$.  
Indeed  for  model 1, all the correlation functions can be computed analytically~\cite{Gaspard}. In particular $< \theta_t \, ; \theta_0> =  2^{-t} / 12 $.    

Now we can also compute the susceptibility  given by the Fourier transform of\rfq{RepCo}.  This gives:
\begin{equation}
\hat{\chi} (\omega)  = \frac{1}{1 - e^{i(\omega + i \log 2)}} =   \frac{2(2-\cos \omega + i  \sin \omega)}{5-4 \cos \omega} 
\label{eq:suscep2}
\end{equation}
Let us remark that $\hat{\chi} (\omega)$ has a complex pole in $\omega = -i \log 2$ on the imaginary axis.   

\subsection{Numerical methods}
\label{sec:method}

The two elementary analytical results\rfq{RepCo}-(\ref{eq:suscep2})  can be used to test numerical methods to compute $\chi(t)$ or $\hat{\chi}(\omega)$.  
In particular it can serve   to validate  the procedure we have used in analysing  the linear response of a chaotic network of interconnected units~\cite{CS1}.

Let us first consider a direct method to compute the impulse response $\chi(t)$, introduced by Falcioni {\em et al}  in~\cite{FIV}.  This method was  proposed   for computing the impulse response to  perturbations independent of the state, but  it is easily adapted  here where  the perturbation depends on the state.
In what follows we discuss the method for the  system $\theta_{t+1} = f(\theta_t)$ introduced above, but of course the method is general.
As before the goal is to compute $<\delta A>_t $ to first order in $\eps$, when the perturbed system follows the same dynamics as the unperturbed one, except that  at time $t=0$ the state  is displaced by $\tilde{\theta}_0 = \theta_0 + \eps X (\theta_0)$. 
 
The principle of the method studied  in~\cite{FIV}  can be described as a finite-sampling estimation of the integral defined by eq.\rfq{prop2}.  
  More precisely, for $t>0$ fixed, $\chi(t)$ is  estimated  by averaging many realisations of $A(f^t(\tilde{\theta}_k))$  ($k=1,\cdots,N$),  starting from  initial data for $\theta_k$ chosen with the SRB measure on the (unperturbed) attractor.  Adapted to the present notations,  the authors of~\cite{FIV}  propose to  evaluated  $\chi(t)$ as follows:
\begin{equation}
\chi(t) \approx  \frac{1}{N\, \eps} \sum_{k=0}^{N} \left(  A \left(f^t[\theta_k + \eps X(\theta_k)]\right) - A\left(f^t[\theta_k ]\right)  \right)
\label{eq:FIV}
\end{equation}
In practise,  reckoning with the ergodicity of one trajectory of the unperturbed dynamics, the set $\{\theta_1, \theta_2,\cdots,\theta_N \}$ can be prepared  by considering only one (long) trajectory $0 < t < T$ and dividing $T$ in $N$  samples, $T= N \tau$, where $\tau$ is  the maximal time for which $\chi(t)$ will be caculated. Then 
 the initial data $\theta_k$   can be simply  chosen as  $f^{\tau} (\theta_{k-1})$.
 
\begin{figure}
\begin{center}
\includegraphics[width=10cm]{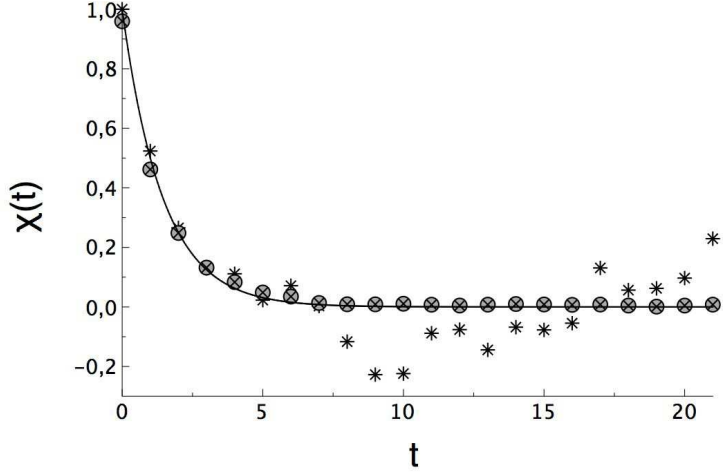}
\caption{ The  impulse response of system\rfq{fperturb}-(\ref{eq:Xtheta})  is given by $\chi(t) = 2^{-t}$  (solid line).   It is compared with  numerical  simulations using  the  estimate\rfq{FIV}  (with $\eps = 0.01,  N = 4552000$, symbol ``{\sf *}''  ).  In this example, it is seen that  large fluctuations  grow  from $t=8$.   The other points ( symbol ``{\sf x}'')  are obtained by using the indirect method which consists in calculating the  (inverse) Fourier transform of estimate\rfq{chiErgoMeth}  (See also Fig.\ref{fig:mod2suscep}).  }
\label{fig:mod2}
\end{center}
\end{figure}

 Now, considering the concrete example for $X(\theta)$  worked out in last Section,
figure~\ref{fig:mod2}   compares  the response function calculated numerically with this method, with the analytical result of eq.\rfq{RepCo}.  It shows a good agreement of the numerics for relatively short times, beyond which error fluctuations increase dramatically.     
In ref.~\cite{FIV} the authors discuss the growth  of the error due to finite size sampling and show that it grows faster than $\exp (\lambda t) / \sqrt{N}$ where $\lambda$ is the largest Lyapunov exponent (here $\lambda = \log 2$).  This difficulty is enhanced in   computing  the susceptibilty $\hat{\chi}(\omega)$, as this  Fourier transform needs a long time estimate of\rfq{FIV}.  
 Therefore in~\cite{CS1}  we have developed a new method which takes the other way round.  We  compute numerically the susceptibility and, if needed, deduce from it the impulse response.

The principle of the new procedure we have implemented  is quite simple~\cite{RuPriv}:   returning to the system\rfq{fperturb}, we know  that the average $<A>_t$ differs from $<A>=\pi$ by $<\delta A>_t = \eps \hat{\chi}(\omega) e^{-i \omega t}$. 
So $\hat{\chi}(\omega) = <\delta A>_t e^{i \omega t}/\eps$ is independent of $t$ and we can write for arbitrary large $T$:
\begin{equation}
\hat{\chi}(\omega) = \frac{1}{T\eps }   \sum_{t=1}^T  <\delta A>_t e^{i \omega t}
\label{eq:chiMeth}
\end{equation}
Let us remark that  in this equation   $<\delta A>_t  =  <A>_t - <A>$   represents  a complex number
because  the  perturbed dynamics given by\rfq{fperturb} is written by means of  the complex perturbation $\eps \xi(\theta)  e^{-i\omega t}$.  In practise, however, i.e. for numerical purpose,   $<A>_t $ can be decomposed in real and imaginary parts,  $<A>_t  =  <A_1>_t  + i <A_2>_t $,  and both of them can be computed separately by considering respectively perturbation of the forms $\eps \xi(\theta)  \cos(\omega t)$ for $<A_1>_t  $ and of the form $-\eps \xi(\theta)  \sin (\omega t)$ for $<A_2>_t  $. 
Now the important point is that the above average $< \; >_t$ could  be attained also by making use of an ergodic assumption.  In other words, for a fixed $\omega$ the time-periodic SRB measure underlying the average $< \; >_t$ 
can be created by  the long time evolution of the perturbed system. Therefore we can estimate\rfq{chiMeth} by writing, for $\omega \neq 0$, small $\eps$ and sufficiently long $T$:
\begin{equation}
\hat{\chi}(\omega) \approx  \frac{1}{T\eps }   \sum_{t=1}^T   A(\tilde{f}^t(\theta))  e^{i \omega t}
\label{eq:chiErgoMeth}
\end{equation}
where $\tilde{f}$ is the perturbed dynamics given by\rfq{fperturb}, modulo its  practical implementation in terms of real and imaginary parts.  Let us notice that in the latter sum we can replace $\delta A$ by $A$ because  the contribution of the non-perturbed dynamics in this
average disappears (for $\omega=0$).  
 
\begin{figure}
\begin{center}
\includegraphics[width=10cm]{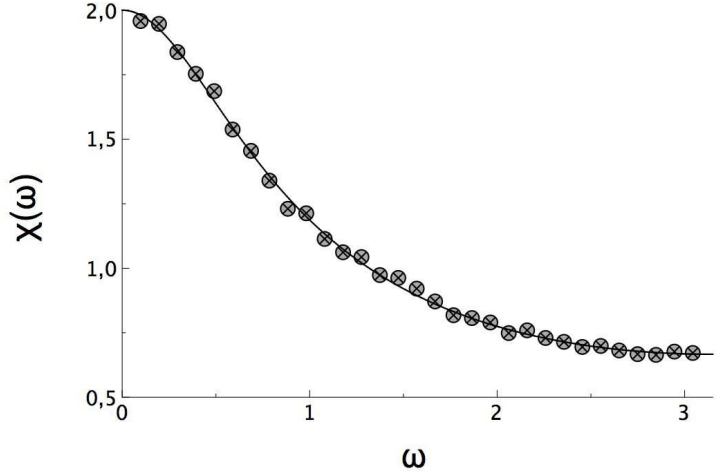}
\caption{Modulus of the complex susceptibility as computed analytically [from eq.\rfq{suscep2}, solid line] and numerically [symbol ``{\sf x}"].  The latter is given by the estimate\rfq{chiErgoMeth}  averaged over $200$ samples  obtained with $\eps= 0.01$ and  $T=2^{19}$. }
\label{fig:mod2suscep}
\end{center}
\end{figure}

Our numerical methods allows one, in principle, to compute the susceptibility $\hat{\chi} (\omega)$ for an arbitrary sampling of $\omega = k 2\pi/M  \in [0,2\pi]$.    The subsequent inverse Fourier transform can then provide the impulse response on the time interval $[0,M]$ with large $M$.  
This procedure has been used with success in~\cite{CS1,CS2}.  Here we illustrate it on the basic example considered above. 
Figure~\ref{fig:mod2suscep} shows comparison of the analytical result of eq.\rfq{suscep2} with the numerically computed susceptibility, which is quite  satisfactory. 
Note that both approximation relations\rfq{FIV}-(\ref{eq:chiErgoMeth}) do not involve an explicit  linearisation (or a differentiation) of the dynamics.   Therefore these relations can be used to numerically tackle the nonlinear response of the system to perturbations.  From the operational point of view, the linear regime is characterised by a non-dependence of the results  in $\eps$.

NB: To obtain these numerical results we had to iterate the dynamics $\theta_{t+1}=2\theta_t \mod 2\pi$ and
it is well known that a naive simulation of this map eventually drives all trajectories to $0$.
We have circumvented the problem by encoding numbers with an (arbitrary long) sequence of bytes
and used the left shift $<<$ available in $C$ language. This shift  set the last bit to zero.
Then we replace it by a bit $0$ or $1$ randomly chosen with a probability $p$ or $q$. For
$p=q=\frac{1}{2}$ we obtain then trajectories typical for the SRB measure. Note that
we can obtain trajectories typical for any Bernoulli measure by changing the probabilities
$p,q$. We believe that this method is well known, though we haven't been able to find
any reference dealing with it.  

\subsection{Dealing with the non-invertibility of $f$}

Now we come back to the general case where the perturbation  $\xi(\theta)$ cannot  be written as $(X\circ f)(\theta)$.    Then,  the change of variable  $\phi = f(\theta)$ used in the integral\rfq{RepContMeas},  can still be performed, provided that the integration domain  $[0,2\pi]$ is decomposed into the subintervals $[0,\pi]$ and $[\pi,2\pi]$ where $f$ is invertible.   In this way the same formal expression\rfq{RepCorrel0} for $\chi(t)$ is obtained as before, but where: 
\begin{equation}
X(\theta) = \frac{1}{2} \left[ \xi(\frac{\theta}{2}) +\xi(\frac{\theta}{2}+\pi)\right]
\label{eq:Xeff}
\end{equation}

 Thus in the chosen example, the effective perturbation $X(\theta)$ appears as the perturbation $\xi$ averaged over the pre-images of $f$.   This concept generalises to general one-dimensional map $f$ defined on an interval $I = \bigcup_k I_k $ assuming the restrictions of $f$ to each $I_k$ invertible    (say $f_k$)   by writing :
\[
X(\theta) = \sum_k  \frac{\xi_k (\theta) \eta_k (\theta)}{| f^{\prime} (f_e^{-k}(\theta)|} \chi_{I_k}((f_e^{-k}(\theta))
\]
where $\chi_{I_k}(\theta)$ is the indicatrix of interval $I_k$;   $\xi_k = \xi \circ f_e^{-k}$ and $\eta_k = \eta \circ f_e^{-k}$,  assuming that the SRB measure is written $\rho_f (d\theta) = \eta(\theta) d\theta$. 
Again a concrete example is easy to work out in the case of model 1.   Let us choose $\xi(\theta) = \theta (2\pi - \theta)$.  Then  eq.\rfq{Xeff} becomes $X(\theta) = \theta(2\pi - \theta)/2 - 1/4$ and it is found that $\chi(t)$  exactly as eq.\rfq{RepCo} but divided by $4$.

\section{Chaotic rotator in the plane}
\label{sec:rotator}

In the previous Section we have seen that in the case of  systems with expanding dynamics only the fluctuation-response theorem applies: under suitable hypothesis,  the response functions are matched with correlation functions.
In view of the simple examples treated above, the technical point which makes the result work  is integrating  by part in eq.\rfq{ImpulLinResp}  (and in example eq.\rfq{RepContMeas}).  In the general case  this integration by part  is no longer possible
(in the stable directions)
 for hyperbolic dynamics with contraction in phase space.  This is because the SRB measure becomes usually singular in the stable directions  so that  it cannot be written with a density function, e.g. $\rho_F(d\bx) \neq  \eta(\bx) d\bx$ for some regular function $\eta(\bx) $.
Nevertheless the response function given  by the integral\rfq{ImpulLinResp}, and so the susceptibility,  can be well defined, but in general cannot  be identified anymore with a correlation function. To analyse this situation, Ruelle decomposes the perturbation giving rise to the response functions  in two parts, by locally projecting  this perturbation respectively onto the stable and the unstable directions of the tangent dynamics.  This can be achieved  when the system is uniformly hyperbolic, i.e. when the tangent space at any point $\bx$  decomposes uniformly with respect to $\bx$  into a direct sum of a stable and an unstable subspace~(e.g. \cite{Gaspard}).

\begin{figure}[htb] 
\centerline {
\includegraphics[width=8cm]{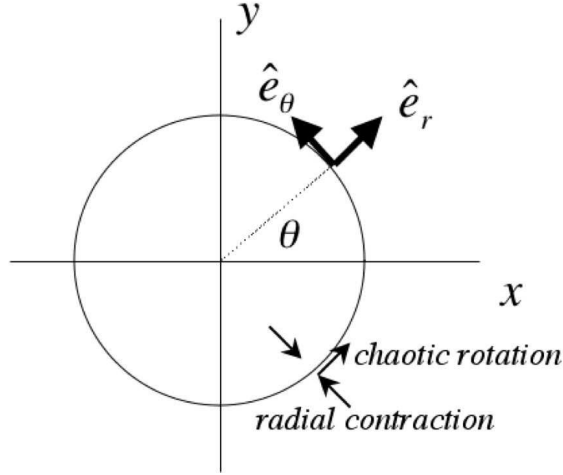}
}
\caption{ \label{fig:phaseportrait} {\protect\small Schematic representation of the phase portrait of the toy model.  The phase plane is shrinked by the dynamics on the unit circle playing the role of the chaotic attractor. Models implementing this scheme are given by eqs.\rfq{trmodel}.  }
}
\end{figure}

The goal of this Section is to illustrate this  decomposition  on  simple models where the uniform hyperbolicity holds by construction and where explicit computations can be done.   To this purpose we consider a class of two-dimensional hyperbolic systems with a  chaotic attractor on  the unit circle, as  represented on figure~\ref{fig:phaseportrait}.  Here the phase space is represented by the plane  which  is contracted  by the dynamics onto   the unit circle  where  there is a rotation with expanding dynamics.  This  might be one of  the simplest example representing a chaotic (non-fractal) attractor.
There are several ways to implement this qualitative dynamics, the simplest one being to decouple the rotation and the radial contraction.  In all the forthcoming implementations  of this class of model, one assumes  that the SRB measure can be factorised as a product of the form:
\[
\rho_F(dxdy) = \rho_F(rdrd\theta)=   \delta(r-1) r dr\,  \eta(\theta) d\theta/2\pi
\]
Here $\delta(r-1)$ is the singular part of the measure, due to the contraction of the radial direction.  On the other hand the measure on the unit circle (which represents the  unstable manifold of any point on the attractor)  is continuous and described by the density function $\eta(\theta)$. The latter  is equal to $1$ in  the particular case of the dynamics $\theta \mapsto 2\theta$  considered in the previous Section.  
Before specifying some implementations of the dynamics of Fig.~\ref{fig:phaseportrait}, we summarize  the objective  of the linear response theory and  introduce  the notations  as follows.  Consider the 2-dimensional time-discrete systems:
\begin{eqnarray}
x_{t+ 1}  & = &  F_x(x_t,y_t) + \eps_x \, \xi_x (t+1,x_t,y_t)  \nonumber \\
y_{t+ 1}  & = &   F_y(x_t,y_t) + \eps_y \, \xi_y (t+1,x_t,y_t) \label{eq:eqxy} 
\end{eqnarray}
where the first term  $\bF = (F_x,F_y)$ is the (autonomous) dynamics which is perturbed by a  time-dependent vector field  of small amplitude ($\eps = |(\eps_x, \eps_y)|  \ll  1$).
Here the coordinates $(x,y)$ do not  necessarily mean  the cartesian coordinates, but general coordinates on the plane.  In the following, we will consider only two cases, however, namely the cartesian or  the polar coordinates.   Furthermore, the  observables  will be the coordinates themselves, except for the ``angle''  variable of the polar coordinates which will be chosen like previously as  $A(\theta)$, i.e. the projection of $\theta$ in $[0,2\pi)$.  

Now suppose that in absence of perturbation the mean values  of $x$ and $y$ are zero.
Then, when the perturbation is turned on,   we are interested to the mean values of  $(x_t,y_t)$  to  first order in $\eps$.    When   $\bxi(t,x,y) = \bxi(x,y) e^{-i \omega t}$, the  solution of this  problem has been discussed above and it    can be written in terms of the susceptibility matrix, namely:
\begin{equation}
\left(\begin{array}{c} 
<\delta x > _t \\
<\delta y > _t
\end{array}\right)   = 
\left(\begin{array}{cc} 
\hat{\chi}_{xx}(\omega)  & \hat{\chi}_{xy}(\omega)\\
\hat{\chi}_{yx}(\omega)  & \hat{\chi}_{yy}(\omega)
\end{array}\right) 
\left(\begin{array}{c} 
\epsilon_x \\
\epsilon_y 
\end{array}\right)  e^{-i \omega t}
\label{eq:xy}
\end{equation}
The  matrix elements $\hat{\chi}_{ij}(\omega) $ ($i,j = x$ or $y$) are the Fourier transform of the impulse response matrix elements  given by eq.\rfq{ImpulLinResp} :
\begin{eqnarray}
\chi_{ij}(t)  & = & \int \rho _F (dxdy)   \nabla F_i ^t  (F(\bx)) \cdot ( \xi_j (\bx) \be_j ) \nonumber \\
& = &  \int \rho _F (dxdy)   \nabla F_i ^t  (\bx) \cdot  ( X_j (\bx) \be_j )  
\label{eq:chi_ij}
\end{eqnarray}
In the last equation  the  function $X_j $  coincides with $ \xi_j \circ \bF^{-1}$  if  the dynamics is invertible.  If  $\bF$ is not invertible then,  as discussed in the previous Section, $X_j$  can be interpreted as  a type  of averaging\footnote{\label{X} In the non-invertible case, a particular case where  the computation of this averaging over the pre-images of $\bF$  gives a well controlable  result is when  the perturbation function is chosen under the form  $\xi_j = X_j \circ \bF$.  Then the result is simply  $X_j$. }   of $\xi_j$ over the pre-images of $\bF$. 

Finally, assuming that the  system is uniformly hyperbolic, i.e. when the tangent space at $\bx$  decomposes uniformly
into a direct sum of a stable and an unstable subspace, the vector $X_j (\bx) \be_j $ appearing 
in\rfq{chi_ij}  will  be decomposed as
\begin{equation}
\label{eq:Xdecomp}
X_j (\bx) \be_j   = \bX_j ^{s} +\bX_j ^{u}
\end{equation}
Then, by substituting this sum in eq.\rfq{chi_ij}  we will be mostly interested in 
 the ensuing  decomposition of $\chi_{ij}  (t)$ into the stable and unstable susceptibilities as follows: 
\begin{eqnarray}
\chi_{ij}  (t) & = &   \chi_{ij}^{(s)} (t) +  \chi_{ij}^{(u)} (t) \label{eq:decomp}
\end{eqnarray}
for $t \geq 0$  (and $\chi_{ij} = 0$ otherwise).

In what follows we start by considering the cases where by construction $\bX_j ^{u}=0$ or $\bX_j ^{s}=0$ in eq.\rfq{Xdecomp}.  This enables one to concentrate on the stable or on the unstable susceptibilities and their  resonances, the former being  the novel feature of out of equilibrium dissipative systems.   

\subsection{Stable resonances}
\label{trmodel}

As a first realisation of the phase portrait depicted on Fig.~\ref{fig:phaseportrait}, we use polar coordinates (i.e. $x \equiv \theta$ and $y \equiv r$ in the eqs.\rfq{eqxy}) and we consider the following system:
\begin{eqnarray}
\theta_{t+ 1}  & = &  g(\theta_t,r_t)  + \eps_{\theta}  \, \xi_{\theta}  (t+1,\theta_t,r_t)  \nonumber  \\
r_{t+ 1}  & = &   R(r_t) + \eps_r \, \xi_r (t+1,\theta_t,r_t)  \label{eq:trmodel}
\end{eqnarray}
with $g(\theta) = 2\theta  + b(r-1)$  ($b$ constant) and $R(r) = 1 + e^{-\mu}  (r-1)$, with $\mu > 0$.

We first  rule out  the case $b=0$.  Then the angular dynamics  is the same as  before, $g(\theta) =  2 \theta$, and the system becomes extremely  simple because the radial and the phase dynamics are decoupled.  The radial dynamics is trivially contracting on the unit circle, with  $r_t = R^t(r_0) =   1 + e^{-\mu t}  (r_0-1)$.  Also,  the impulse response  matrix is diagonal, because  $\hat{\chi}_{r\theta} (\omega) = \hat{\chi}_{\theta r} (\omega) =0$ which is readily deduced from eq.\rfq{chi_ij}.   Using the same equation, the diagonal elements are computed as follows:
\begin{eqnarray}
\chi_{\theta\theta} (t)  & =  & \int \! \rho_F(rdrd\theta)  \,  \partial_{\theta}  f^t (\theta) \, X_{\theta} (\theta,r) = -  < \theta_t \, ; \partial_{\theta} X_{\theta} (\theta_0,1) >
\label{eq:RepCorrel2} \\
\chi_{rr} (t)  & =  & \int \! \rho_F(rdrd\theta)  \,  \partial_{r}  R^t (r) \, X_{r} (\theta,r) = 
e^{-\mu t}  <  X_r (1,\theta) >
\label{eq:RepR}
\end{eqnarray}
for $t \geq 0$. 
The first line is identical to the response function\rfq{RepCorrel1} obtained for the basic model studied in the preceding Section.  This  unstable response function, corresponding to perturbation parallel to the attractor,  is thus a correlation function.  As  an illustrative example we can calculate these integrals 
in the case of perturbation function  $\xi_r (t,\theta,r) =\xi_{\theta}  (t,\theta,r) = X(\theta) \delta_{t0}$, where 
$X(\theta)$  is chosen as  before (cf.  eq.\rfq{Xtheta}).  Then  the susceptibility $\hat{\chi}_{\theta\theta} (\omega)$ is given    explicitly by  eq.\rfq{suscep2}.  One notices  that  it possesses a pole\footnote{
More generally, the Ruelle-Pollicott resonances of this map are given by $-ik \log 2$ where $k$ is a positive
integer \cite{Gaspard}.}   in $\omega = -i \log 2$.
On the other hand, eq.\rfq{RepR}  is an example of stable response,   i.e. the    response to a perturbation along  the contracting  direction of  the attractor.    Note that the result is calculated without integration by part.  The corresponding susceptibility is then computed   as:
\begin{equation}
\hat{\chi}_{rr} (\omega)  = \frac{C}{1 - e^{i(\omega + i\mu)}} 
\label{eq:suscepr}
\end{equation}
where $C$ is a constant. Now this function has a pole located on the imaginary axis  in $\omega = -i\mu$.  This pole is independent of the ones of the correlation function. In particular it can be situated arbitrarily close to the real axis.
This simple example clearly shows the effect of two types of perturbation on the system, either transverse or parallel to the attractor.

Let us deal with the case $b\neq 0$, which introduces  a skew dynamics between the stable and the unstable variables.   Then, if one starts with an initial condition  $(\theta_0,r_0)$ the angular dynamics is easily worked out, giving the  following expression : 
\begin{equation}
\theta_t = g^t(\theta,r) = \left(  2^t \theta + b(r-1) \frac{2^t-e^{-\mu t}}{2-e^{-\mu}}  \right) \: (\mod 2 \pi)
\label{eq:gtr}
\end{equation}
Let us notice that  the dynamics {\em on} the attractor ($r=1$) is  the same as before. This implies that the diagonal terms of the susceptibility remains the same as for $b=0$, and  the correlation functions do not change.  However, the  dependence of $g^t$  on $r$ creates now a  non-diagonal term  which can be written as follows:
\begin{equation}
\chi_{\theta r} (t) = b \frac{2^{t}-e^{-\mu t}}{2-e^{-\mu}} \,  B(t)   \quad \quad \mbox{\rm with} \quad 
B(t) =   \int_0^{2\pi} \! X(\theta) \, d\theta  - \sum_{k=1}^{2^t}  X(  k  \frac{2\pi}{2^{t}}) 2^{-t}
\label{eq:chitr1} 
\end{equation}
 Let us remark that the  factor  $B(t)$  can be interpreted as an ``error''  function measuring   the difference between the integral of $X(\theta)$
 and its approximation by a Riemann  sum  with intervals $2^{-t}$.  (This factor $B(t)$  stems from  the distribution derivative of the ``angle'' observable 
eq.\rfq{distribution} ).   
It is clear that if $X(\theta)$ is Riemann integrable, then the ``error  $B(t)$ has to decrease to zero  when $t$ becomes large.   Actually,  in view of eq.\rfq{chitr1}  it should decrease faster than $2^{-t}$  to get $ \chi_{\theta r} (t)$ going to $0$ with $t$ large. 
For example in the case of $X(\theta)$ given by eq.\rfq{Xtheta},  we obtain  $B(t) = 2^{-2t}$ and thus 
\begin{equation}
\chi_{\theta r} (t) = b \frac{2^{-t}-e^{-(2 \log 2+\mu)  t}}{2-e^{-\mu}} 
\label{eq:chitr}
\end{equation}
This geometrical interpretation of function $B(t)$  shows directly  that if $X(\theta)$  varies much in $[0,2\pi]$ then it will take a long time $t$ to reach  an actual exponential decrease of $B(t)$.
By the meantime the response function can transiently  grow a lot due to the exponential factor $2^t$.  Therefore the degree of variation of $X(\theta)$  determines  the transient growth of the stable response function.  This property will be used   later in Section~\ref{sec:GenCoord} by considering  a perturbation function of the form $X(\theta) = \cos (2^p \theta)$.

 

Now, the susceptibility $\hat{\chi}_{\theta r}(\omega) $, which can be readily deduced from\rfq{chitr}, possesses  a pole situated in $\omega =  - i( 2 \log 2 + \mu)$ on the imaginary axis.   The $\mu$ contribution  is clearly due to the contracting dynamics, whereas the first term $2 \log 2 $ is  in fact a  Ruelle-Policott resonance  (in the present case  they are  multiples  of $\log 2$). 
Thus this example shows a new feature which can appears when the variables are not trivially decoupled into stable and unstable variables:   some stable resonances become  a (linear) combination of poles due to the contracting direction and of resonances  belonging to  the correlation functions.  An attempt of generalisation of this property is given  in the Appendix.

 \subsection{Stable resonances with arbitrary frequencies} 
 \label{sec:network}
   
 The elementary model described by eq.\rfq{trmodel}  has only a one-dimensional radial dynamics.  Consequently any stable pole of the corresponding susceptibility  will be located along  the imaginary axis. 
 On the other hand it is possible  to generalise this model in order to show that stable poles can have  arbitrary frequencies and relaxation times.  It suffices to increase to number of degrees of freedom of the system.  To illustrate this point we present in this section two examples  for which the stable resonances can have arbitrary non zero frequencies.
 
 A simple way to add more degrees of freedom  to the radial dynamics is to transform  its dynamics  into a higher order reccurence. For instance let us consider a  new type of  rotator whose equations  are  a  variation of system\rfq{trmodel}  taking  the form:
 \begin{eqnarray}
\theta_{t+ 1}  & = &  g(\theta_t,r_t)    \nonumber  \\
r_{t+ 1}  & = &   R(r_t,r_{t-1})   \label{eq:trrmodel}
\end{eqnarray}
with as before  $g(\theta) = 2\theta  + b(r-1)$  but  now
$R(r,\varrho) = 1 + 2 e^{-\mu} \cos(\omega_0) (r-1) - e^{-2\mu} (\varrho - 1) $   ($\mu >0$).
The choice of this dynamics  is motivated by the fact that it  creates  in the stable susceptibility  a  pole placed  simply at  $ \omega_0 -i\mu $.    Thus we can easily control its frequency by varying  $\omega_0$.   This example is easy to deal with because the radial dynamics can be solved exactly, and  it can be checked that  given  $r_0$ arbitrary  and $r_{-1} = 0$,  one gets:
\begin{equation}
r_t = 1 + (r_0 -1) e^{-\mu t} \sin \omega_0 (t+1) / \sin \omega_0
\label{eq:r_t}
\end{equation}
     This case  will be considered  below in cartesian coordinates.

 Another way to increase the number of degrees of freedom is to consider a {\em network} of interconnected  rotators.  This example is   interesting  because it  has links with the chaotic  network model that we have  studied previously~\cite{CS1,CS2} and which is mentionned  in the introduction of  this paper.
 So we consider now a collection  of $N$  interconnected  rotators  of the same type as before (cf.  fig.~\ref{fig:phaseportrait}),  assuming for simplicity that the coupling between the units occurs only by means of  their radial variables.  The equations  of this model are the following:
\begin{eqnarray}
\theta_{i, t+ 1}  & = &  g(\theta_{i,t},r_{i,t})  
  \nonumber  \\
r_{i, t+ 1}  & = &   1 +  \sum_{j=1}^N J_{ij} (r_{j,t} - 1)
 \label{eq:trNetModel}
\end{eqnarray}
where $i \in \{ 1, \cdots, N\}$ denote the index of the i$^{th}$ rotator in the network.   The function $g(\theta,r) = 2\theta + b(r-1)$ is the same as before and the real numbers  $J_{ij}$  form  a matrix $\mbox{ \sf  J}$.  The only requirement about  this matrix  is that its spectrum lies  inside the unit circle such that it  gives  in fact a contraction of the radial variables onto $r_{i} = 1$ for any $i$.      On the other hand  we can interpret  $\mbox{ \sf  J}$    as the  {\em connectivity}  matrix of the network, meaning  that the rotator $j$  influences rotator $i$ iff  $J_{ij} \neq 0$.  Moreover this interaction is signed, with the interpretation that positive  $J_{ij}$  means activation and inhibition is described by negative $J_{ij}$.


The model\rfq{trNetModel}   is a straightforward  generalisation of system\rfq{trmodel}, so that the explicit time evolution for $(\theta_{i,t},r_{i,t})  $ can also be calculated.   Furthermore,  if  $\mbox{ \sf  J}$ is diagonalisable, with  $\mbox{ \sf  P}^{-1}\mbox{ \sf  J}\mbox{ \sf  P}= diag\{ e^{-\mu_1 +i \omega_1}  , \cdots, e^{-\mu_N +i \omega_N}  \}$ , then 
the stable  complex susceptibilities (with respect to a perturbation of amplitude $\eps X(\theta)$  in the $r_j$ direction  as considered  in the single rotator model)   can be expressed as follows:
\begin{eqnarray}
\hat{\chi}_{r_i r_j } (\omega) & = &  \sum_k \frac{P_{ik} P^{-1}_{kj}}{1-e^{i(\omega -  \omega_k ) -  \mu_k}} \nonumber  \\
\hat{\chi}_{\theta_i r_j } (\omega) & = &  \sum_k P_{ik} P^{-1}_{kj}
( \frac{  1}{1- e^{i\omega-\log 2 } } -  \frac{ 1 }{1-e^{i(\omega -  \omega_k) - (2\log 2 + \mu_k)}} )
\label{eq:chirr}
\end{eqnarray}
So, it is readily seen that  there are  stable poles  situated in the lower half-complex plane  in  $\omega= \omega_k - i \mu_k  $  and in $\omega= \omega_k - i (2\log 2 +\mu_k) $,  each time $e^{-\mu_k +i \omega_k}$ is an eigenvalue of matrix $\mbox{ \sf  J}$. 
Therefore, although in this model  the unstable resonances are located only on the imaginary axis,  depending on the connectivity matrix there can be an arbitrary number of stable resonances with non-zero frequencies $\omega_k$.     The latter will  not be visible in the power spectrum of the chaotic dynamics, but  eqs.\rfq{chirr}  shows that  a  periodic forcing  of unit  $j$ at  frequency $\omega_k$ can produce a resonance in the linear response of unit $i$.   As discussed in ref.~\cite{CS2},  this property can be used to transmit a signal from $j$ to $i$  (possibly with amplitude modulation).   Moreover we see that   a necessary condition to make this transmision possible is that    $ P_{ik} P^{-1}_{kj} \neq 0$. 
This condition is  satisfied if  the $k^{th}$  eigenmode of $\mbox{ \sf  J}$ has a nonzero component $i$ in the canonical basis and reciprocaly  if the $j^{th}$  vectors of that canonical basis has a nonzero component $k$ when decomposed in the eigenvector basis.
(Roughly speaking, an excitation of the node $j$ excites
the eigenmode $k$, which excites the node $i$).
  This opens the possibility,  for appropriately chosen  connectivity matrices $\mbox{ \sf  J}$, to  get  resonances which are specific to the pair $ij$. Note in particular that if the network as some modular structure such
that the matrix $\sf  P$ has a block diagonal structure then, obviously, there are resonances specific to each 
block. This structure is revealed by our excitation-response procedure even when one has no access to the
explicit form of the dynamics.

 \subsection{Relaxation time longer than the mixing time}
 \label{sec:mixing}
 
 To summarise, in the previous sections we have seen that the linear response between two variables
  $(x_i,x_j)$ of a uniformly hyperbolic  system  can be characterised by  poles  of its susceptibility 
  $\hat{\chi}_{ij} (\omega) $.  A given pole  $\omega = \omega_k - i \mu_k$  of this function is associated with two time scales, namely its resonance frequency $ \omega_k$ and its relaxation time $ \mu_k^{-1}$.  The latter  gives an estimation of the decay time of oscillations $ \omega_k$   when  the  system is submitted to an impulse perturbation via its  $x_j$ variable.  
 Moreover,  in view of the decomposition  $\hat{\chi}_{ij} (\omega)  = \hat{\chi}_{ij}^{(u)} (\omega)  +\hat{\chi}_{ij}^{(s)} (\omega) $  introduced in eq.\rfq{decomp}, there can be  two types of poles.  The  first class of poles are the ones of $\hat{\chi}_{ij}^{(u)} (\omega) $.  They coincide with  the Ruelle-Pollicott resonances, which are also the poles of the power spectrum of a generic variable of the chaotic system.   The knowledge of the closest pole to the real axis  in this class  provides the {\em mixing time} $\tau_m$, defined as the inverse of its imaginary part.   
 This is also the typical decay time of the correlation functions which can be interpreted as the  ``decorrelation'' time   of  state variables of the chaotic attractor.   In the basic examples treated above  $\tau_m = 1/\log 2$.   The second class of poles
 is formed by   the poles of the stable susceptibility $\hat{\chi}_{ij}^{(s)} (\omega) $.  
 In particular their imaginary parts characterise the average decay time of an impulse perturbation along the stable direction, and  thereby the average time to come back  to  ``equilibrium''  after this perturbation.
 For instance  in  the example of the previous Section, eq.\rfq{chirr} shows that  the return time of  $<\theta_i>_t$   to equilibrium      after an  impulse perturbation in the stable direction  $r_j$   is  dominated   by  the largest   $ \tau_{ij} = \max_k (2\log 2 + \mu_k)^{-1}$,   ($k$ chosen in  $\{ 1,\cdots,N$ such that $P_{ik} P^{-1}_{kj} \neq 0 \}$).     Thus  in  this example  $\tau_{mixing}  > \tau_{ij}$  for all pairs $ij$   and this  corroborates the common belief that the return time to equilibrium  of a physical system is given by its mixing time.   
 
 This property is not true in general, however.    The goal of the present Section is to support this statement by providing a generalisation  of our simple model\rfq{trmodel}  where indeed the decay of response functions can be much longer than the mixing time.  Moreover since this property was already observed in our previous study of neural network dynamics~\cite{CS2},   the present example indicates also a possible mechanism to understand this  non-classical  behaviour. 
 
\begin{figure}[htb] 
\centerline {
\includegraphics[width=8cm]{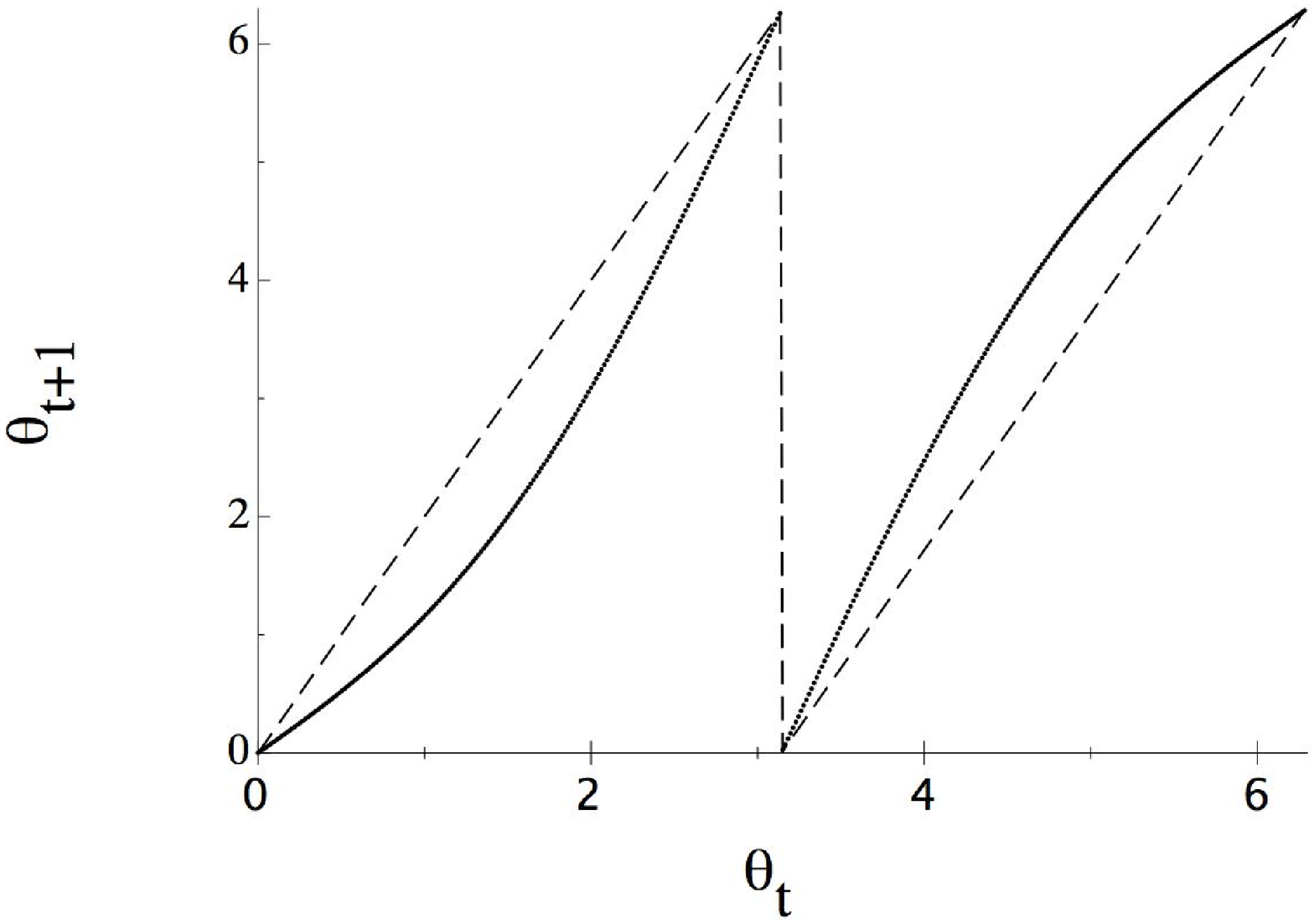}
}
\caption{ \label{ReturnNtheta} {\protect\small
First return map representation of the $\theta$ dynamics defined by eq.\rfq{trsinmodel} in the two limiting case $(r-1) \sim  0$ (dashed line) and $(r-1) \gg \eta  $ (solid line).
}}
\end{figure}

\begin{figure}[!ht]
\vspace{-1cm}
\centerline {
\includegraphics[width=10cm]{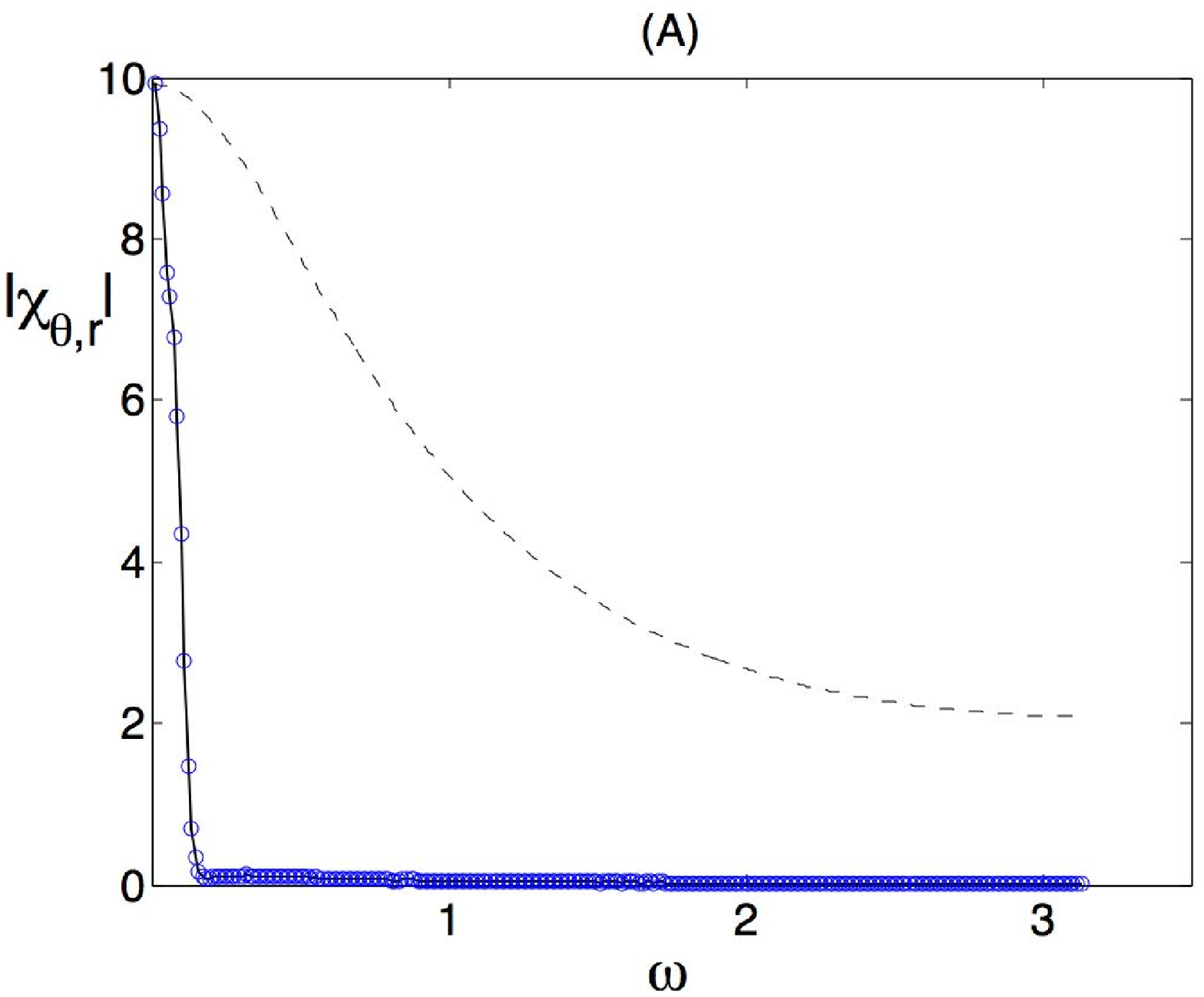}
\includegraphics[width=10cm]{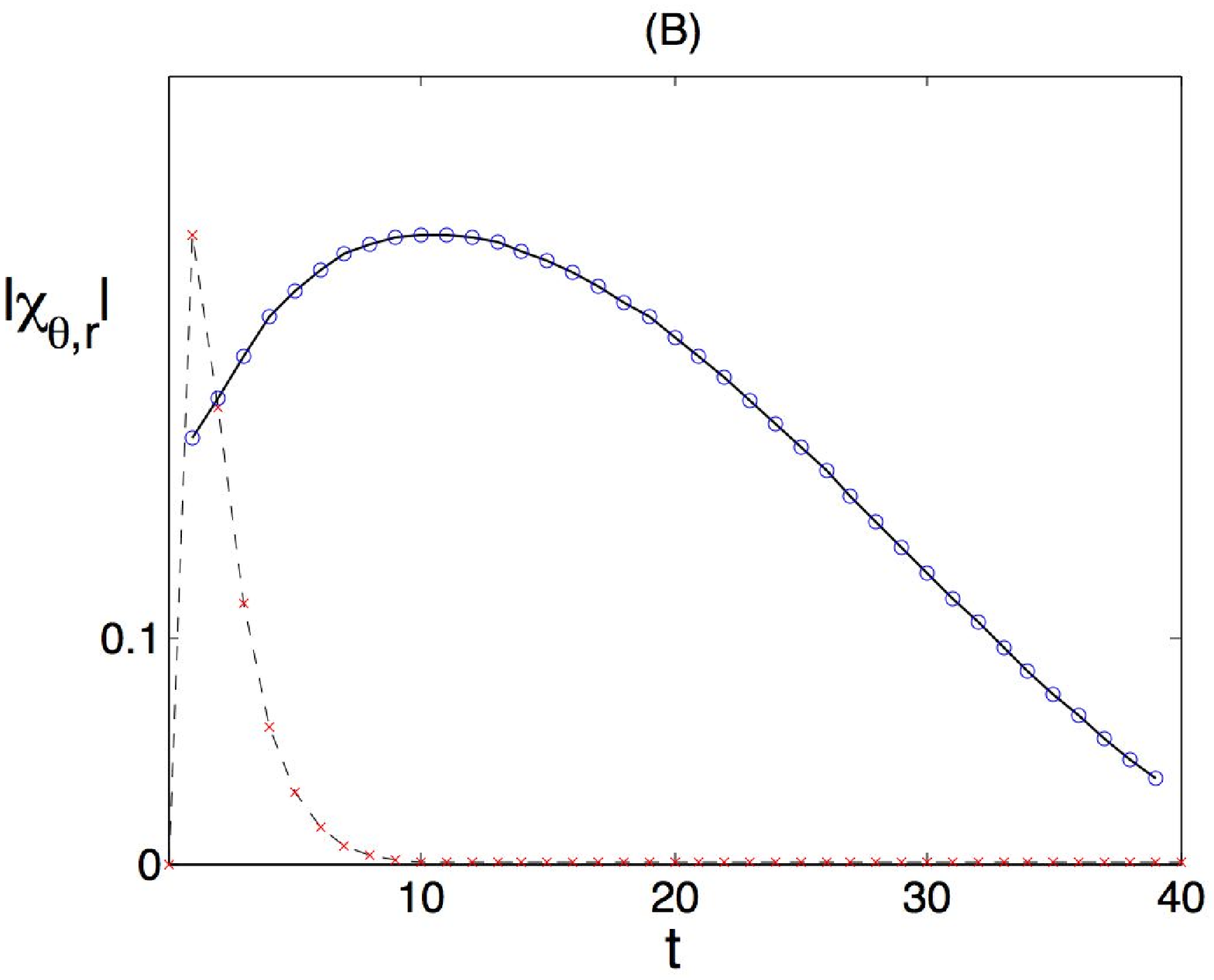}
\vspace{-3cm}
}
\caption{ \label{chitr_om} {\protect\small
(A)  Modulae of the susceptibility $\hat{\chi}_{\theta r} (\omega)$  computed  for the model\rfq{trsinmodel} for different values of $\eta$.  The dashed line curve is the analytical prediction of $|\hat{\chi}_{\theta r} (\omega)|$  in the case $\eta = \infty$.     The case $\eta = 0.05$ (symbol `{\sf o}' )  is numerically computed with estimate\rfq{chiErgoMeth} ($\eps = 0.01, T= 2^{19}$ and yet averaged over $200$ samples).  It  shows a remarkable narrowing of the peak.  (B)  Impulse response functions  $\chi_{\theta r} (t)$ computed  from the inverse Fourier transform of $\hat{\chi}_{\theta r} (\omega)$.   In both cases (A) and (B) the height of curves $\eta = \infty$ have been scale to the same amplitude than $\eta = 0.05$. 
}}
\end{figure}

Let us  consider the following system, which is an  extension of system\rfq{trmodel}:
\begin{eqnarray}
\theta_{t+ 1}  & = &  2\theta_t  + b(r_t) (r_t-1)  + (b(r_t) - 1) \sin \theta_t   \nonumber  \\
r_{t+ 1}  & = &   1 + e^{-\mu}  (r-1)    \label{eq:trsinmodel}
\end{eqnarray}
with  $b(r) =  e^{-(r-1)/\eta}$  and $\eta > 0$.   Thus in the neighborhood of $|r-1| \ll \eta $, the new term $ (b(r_t) - 1) \sin \theta_t$ is negligible and the system then behaves exactly as  the basic model of 
Section~\ref{trmodel}.    This is no longer the case when $ r-1$ increases, entailing that  $b(r)$ decreases to $0$.    Figure~\ref{ReturnNtheta} shows the $\theta$ dynamics in the two limits $r=1$ and $r=\infty$.
This model is hardly  tractable analytically because  it is hard to compute $(\theta_t, r_t)$ as a function of $(\theta_0, r_0)$.    But the dynamics  can still  be analysed by making use of  the numerical method described in Section~\ref{sec:method}.     Figure~\ref{chitr_om}A shows a superposition of the susceptibilities $|\hat{\chi}_{\theta r} (\omega) |$  obtained from two  numerical computations performed with two different values of $\eta$.   For  $\eta = 1$ in eqs.\rfq{trsinmodel}, the numerical results can be  well fitted with the analytical curve computed in the case $\eta = \infty$  (i.e. the Fourier transform of eq.\rfq{chirr} or, equivalently eq.\rfq{chitr} in the case of only one unit).  Then the width of the resonance in $\omega = 0$ is dominated by $ \log 2$ 
corresponding to the inverse of the mixing time.
On the other hand the narrower curve (with symbols ``x'')  in fig.~\ref{chitr_om}  shows an example  of  $|\hat{\chi}_{\theta r} (\omega) |$  computed for small $\eta$.    In this case it is seen that  the width of the resonance in $\omega = 0$  has substantially decreased.   As a matter of fact it can be arbitrarily decreased by diminishing  further $\eta$.   In the same time the height of the resonance peak increases.  (This is not visible on the figure because their maxima have been scaled to the same height).
Thus  the susceptibility  $\hat{\chi}_{\theta r} (\omega)$  diverges with $\eta \rightarrow 0$ because  one of its poles    pinches the real axis in this limit.    The reason for this behavior can be  understood  at least qualitatively as follows.      As $r-1$ increases from $\eta$,  the system\rfq{trsinmodel}  looses indeed  its  uniform hyperbolicity. This is clear  on fig.~\ref{ReturnNtheta},  where it is seen that it   $r-1 \gg \eta$ then  the  first return map for $\theta_t$   acquires a slope  1 in abscissa  $\theta = 0$.   This means that a perturbation along the radial direction  can induce a drastic slowing down of the ``chaotic'' rotation.   The latter is resumed only when the  radial perturbation has sufficiently relaxed.  Thus in this case this is the decay time along the contracting direction which controls the return time to  equilibrium.

Now one interesting physical  consequence of this phenomenon appears by reconstructing the impulse response function $\chi_{\theta r} (t)$ from the inverse Fourier transform of $\hat{\chi}_{\theta r} (\omega)$.   Figure~\ref{chitr_om}B  shows the comparison between the cases $\eta$ large (cf.  eq.\rfq{chitr}) and $\eta$ small.  In the first case the response of $\theta$ decays as expected like $2^{-t}$ which is the typical decay time of the correlation functions of this system. 
In the second  case it is seen that  the system responds with an impulse of large amplitude which moreover decays on a  time of order 10 times longer than the mixing time.    Therefore this example illustrates a novel feature of Ruelle's theory, compared with the common intuition  of   nonequilibrium physical (conservative) systems:  when a dissipative  system is pushed out of equilibrium, its return time to equilibrium can be substantially longer than the mixing time which could be predicted  by  looking at its correlation functions.  This property  merely  reflects the possible violation of the fluctuation-response theorem in dissipative systems. 

The numerical simulations reported in this section concerned our single rotator which is a 2 degrees of freedom system.  Nevertheless we anticipate that the same type of  phenomenon will show up in the network of rotators described by eqs.\rfq{trNetModel}.  In fact this behaviour of having a longer return time to equilibrium than the mixing time    was also observed in the numerical simulations of our neural network dynamics~\cite{CS2}.

\subsection{Decomposing the linear response in general coordinates}
\label{sec:GenCoord}

The model\rfq{trmodel}  is sufficient  to show the existence of two types of responses in hyperbolic systems. However, the choice of variables  $(\theta, r)$  made to perturb this system and to observe its response
 is quite particular because its coincides  with the local unstable and  stable directions  $ \{ \be_{\theta}, \be_{r} \} $.  So the decomposition of the response functions, or of the susceptibilities, into their stable and unstable parts is  trivial for this choice. 
 In this Section, we analyse the same model, but with observables which are the   cartesian coordinates in the plane.   The unperturbed dynamics is thus the same but we change the way to perturb it  and  the variables to look at.  
  The problem becomes slightly more complicated because 
    each element of the the susceptibility matrix have  both components, stable and unstable.  The identification  of these   two contributions  in a response function   can scarcely be done explicitly  in a general  system because the projectors onto the local stable or unstable manifolds are not easy to compute for an arbitrary system.     Nevertheless   the present model is sufficiently simple to perform   this decomposition  analytically  and  to discuss some of its consequences  on the resonances of the susceptibility. 

Let us consider the   model\rfq{trrmodel} introduced  before  but now written  in cartesian coordinates $(x,y)$ of the plane:
\begin{eqnarray}
x_{t+ 1}  & = &  F_x(x_t,y_t,x_{t-1},y_{t-1}) + \eps_x \, \xi_x (t+1,x_t,y_t)  \nonumber \\
y_{t+ 1}  & = &   F_y(x_t,y_t,x_{t-1},y_{t-1}) + \eps_y \, \xi_y (t+1,x_t,y_t) \label{eq:eqxyBIS} 
\end{eqnarray}
with the functions $\bF = (F_x, F_y)$  defined by:
\begin{eqnarray}
F_x(x_t,y_t,x_{t-1},y_{t-1})  & = &  R(\sqrt{x_t^2+y_t^2},\sqrt{x_{t-1}^2+y_{t-1}^2}) \cos ( g[\arg(x,y),\sqrt{x_t^2+y_t^2}])    \nonumber \\ 
F_y(x_t,y_t,x_{t-1},y_{t-1})  & = &   R(\sqrt{x_t^2+y_t^2},\sqrt{x_{t-1}^2+y_{t-1}^2}) \sin ( g[ \arg(x,y),\sqrt{x_t^2+y_t^2}])  
\label{eq:eqxyBISF}
\end{eqnarray}
and  the functions  $R$ and $g$ have already been defined after eq.\rfq{trrmodel}.  The notation   $\arg(x,y) $  means  $\arctan(y/x)$  for $x>0$ and $\arctan(y/x)+\pi$ for $x<0$.   
Thus in the absence of perturbation ($\eps_x = \eps_y = 0$), the standard  change of variable $x = r \cos \theta$, $y= r \sin \theta$ gives the unperturbed model\rfq{trrmodel}.  

Assuming without loss of generality  that the mean values of the variables are zero  in the unperturbed situation,    $<x>=0=<y>$, we are interested in the mean values $<x>_t$ and $ <y>_t$ for small $\eps$ and a prescribed form of the perturbation function $\bxi (x,y)$. As already discussed (see footnote~\ref{X}),  in order to simplify the formalism we choose a perturbation  the form $\bxi = \bX \circ \bF$, with $\bX = X_x \be_x$   or  $\bX  = X_y \be_y$.    The goal is now  to compute the
susceptibility matrix  $\hat{\chi}_{ij}(\omega)$, or equivalently   the corresponding response matrix  ${\chi}_{ij}(t)$  defined in eq.\rfq{chi_ij},  as well as  its decomposition   in  a sum of an unstable  and of a stable part.  
To this purpose, by construction in the present model  the perturbation $X_j \be_j$ ($j=x$ or $y$) can be decomposed along   its stable and its unstable projections, for example :
\begin{equation}
\label{eq:chiij}
X_y  \be_y = X_y  \sin \theta \,  \be_{r} +  X_y  \cos \theta \,  \be_{\theta} 
\end{equation}
where $ \be_{r}$ is the radial unit vector (local stable direction) and $ \be_{\theta}$  is the tangential unit vector  to the circle (local unstable direction). 
%
Thus from eqs.\rfq{chi_ij}-(\ref{eq:decomp}), and by using  the equations  $\nabla F_i . \be _r = \partial_r F_i $ and $\nabla F_i . \be_{\theta} = \frac{1}{r} \partial_{\theta} F_i$  ($i=x$ or $y$) , one can write the decomposition   of $\chi_{xy}  (t)$   into the stable and unstable susceptibilities as follows: 
\begin{eqnarray}
\chi_{xy}  (t) & = &   \chi_{xy}^{(s)} (t) +  \chi_{xy}^{(u)} (t) \label{eq:decomp2} \\
& = &  \int \! \rho_F(d\bx) \, \sin \theta \,  \partial_r F_x^t (\bx) \, X_y (\bx)  +   \int \! \rho_F(d\bx) \,  \frac{\cos \theta}{r} \partial_{\theta} F_x^t  (\bx) \, X_y  (\bx)   \nonumber
\end{eqnarray}
for $t \geq 0$  (and $\chi_{xy} = 0$ otherwise).  The other matrix elements $\chi_{xx}(t) $ or $\chi_{yy}(t) $ have an anologous expression.  Let us recall that the invariant measure to be used in these integrals takes the following  form (in  polar coordinates) :
$ \rho_F (d\bx) =    r  d\theta \,   \delta(r-1)\delta(\varrho-1) dr d\varrho $  
with $\delta$ denoting here the   Dirac  distribution. 
 This enables one to get the following results:
\begin{eqnarray}
\chi_{xx}^{(u)} (t)   & = &  < \cos \theta_t ;\, \partial_{\theta} [\sin \theta_0  X_x(\theta_0 )] > \label{eq:chixxu}    \\
\chi_{xx}^{(s)} (t) & =  &  \partial_r R^t < \cos \theta_t; \, \cos \theta_0 X_x(\theta_0) >  \label{eq:chixxs} \\
\chi_{xy}^{(u)} (t)   & =  & - < \cos \theta_t ;\, \partial_{\theta} [\cos \theta_0  X_x(\theta_0 )] >   
\label{eq:chixyu} \\ 
\chi_{xy}^{(s)} (t)   & =  &  - \partial_r g^t < \sin \theta_t ;\, \sin \theta_0  X_y(\theta_0 ) > \label{eq:chixys} 
\end{eqnarray}
(We do not write the corresponding functions $\chi_{yy}^{(u)} (t)$ and  $\chi_{xx}^{(u)} (t)$ which are quite similar).   We have only considered a $\theta$-dependence of the perturbation function $X_i(\theta)$ (because a $r$ dependence would be trivial due to the Dirac distributions).     As it is expected the unstable response functions\rfq{chixxu} and (\ref{eq:chixyu})  are pure correlation functions, whereas the stable responses\rfq{chixxs} and (\ref{eq:chixys})  are correlation functions multiplied by a factor  related to the  contracting dynamics.   As already discussed this latter term is responsible for creating poles in the susceptibility which are not observable in the powerspectrum.   

Now, another  point that we want to illustrate  in this section is the following:  as mentionned above, the decomposition $ \chi_{ij}^{(s)} (t) +  \chi_{ij}^{(u)} (t)$ can scarcely be done in practise.   In this case the resonances which can be observed  are induced from  the maxima of the
 modulus  $| \chi_{ij} (t)|$.     But  the latter can be misleading to detect actual poles of the  stable or of the unstable susceptibilty.  To make this point clearer, we treat a concrete example. 

Let us consider again  the  perturbation function  $X(\theta) = \theta(2\pi - \theta)$  to be used in eqs.\rfq{chixxu}-(\ref{eq:chixys}) and let us compute  explicitly $\chi_{xx}(t) = \chi_{xx}^{(u)} (t) +  \chi_{xy}^{(s)} (t)$.   
The first term is given by the correlation function\rfq{chixxu} which can be written as:
\begin{equation}
\chi_{xx}^{(u)} (t) =   \int_0^{2\pi} \!  d\theta \, \cos ( 2^t \theta)  \partial_{\theta} [ \theta(2\pi -\theta) \sin \theta ]  =  \left\{\begin{array}{l} \frac{\pi}{2}  + \frac{2}{3} \pi^3    \mbox{\ \ if\ } t=0  \\ 
 -2\pi \left(  \frac{1}{(2^t + 1) ^2 } + \frac{1}{(2^t - 1) ^2 }  +  \frac{2}{(4^t -1)  }   \right)  \mbox{\ \ if\ }  t > 0   \end{array}\right.
\label{eq:chixx^u}
\end{equation}
and, as usual, $\chi_{xx}^{(u)} (t) =0$ for $t < 0$.
Next we compute the stable susceptibility $\chi_{xx}^{(s)} (t) $ by evaluating the two factors of\rfq{chixxs}.  The first factor $\partial_r R^t$   is worked out by using\rfq{r_t} and  one  gets
\begin{equation}
\chi_{xx}^{(s)} (t)  =   \frac{\alpha^t \, \sin \omega_0 (t+1) }{\sin \omega_0 } \,  C(t) 
\label{eq:chixx^s}
\end{equation}
with the second factor being  the correlation function:
\begin{equation}
C(t)  =  \int_0^{2\pi} \!  d\theta \, \cos ( 2^t \theta)\cos( \theta)  \, \theta(1-\theta) =  \left\{\begin{array}{l} -\frac{\pi}{2}   + \frac{2}{3} \pi^3    \mbox{\ \ if\ } t=0  \\ 
 -2\pi\left(  \frac{1}{(2^t + 1) ^2 } + \frac{1}{(2^t - 1) ^2 }      \right)  \mbox{\ \ if\ }  t > 0   \end{array}\right.
\label{eq:Cxx}
\end{equation}
Finally the complex susceptibility is obtained by summing the stable and the unstable contributions.  For example  $\chi_{xx}(t) = \chi_{xx}^{(u)} (t) + \chi_{xx}^{(s)} (t)$ with eqs.\rfq{chixx^u}-(\ref{eq:Cxx}).

\begin{figure}[!ht]
\vspace{- 1cm} 
\centerline {
\includegraphics[width=10cm,clip=false]{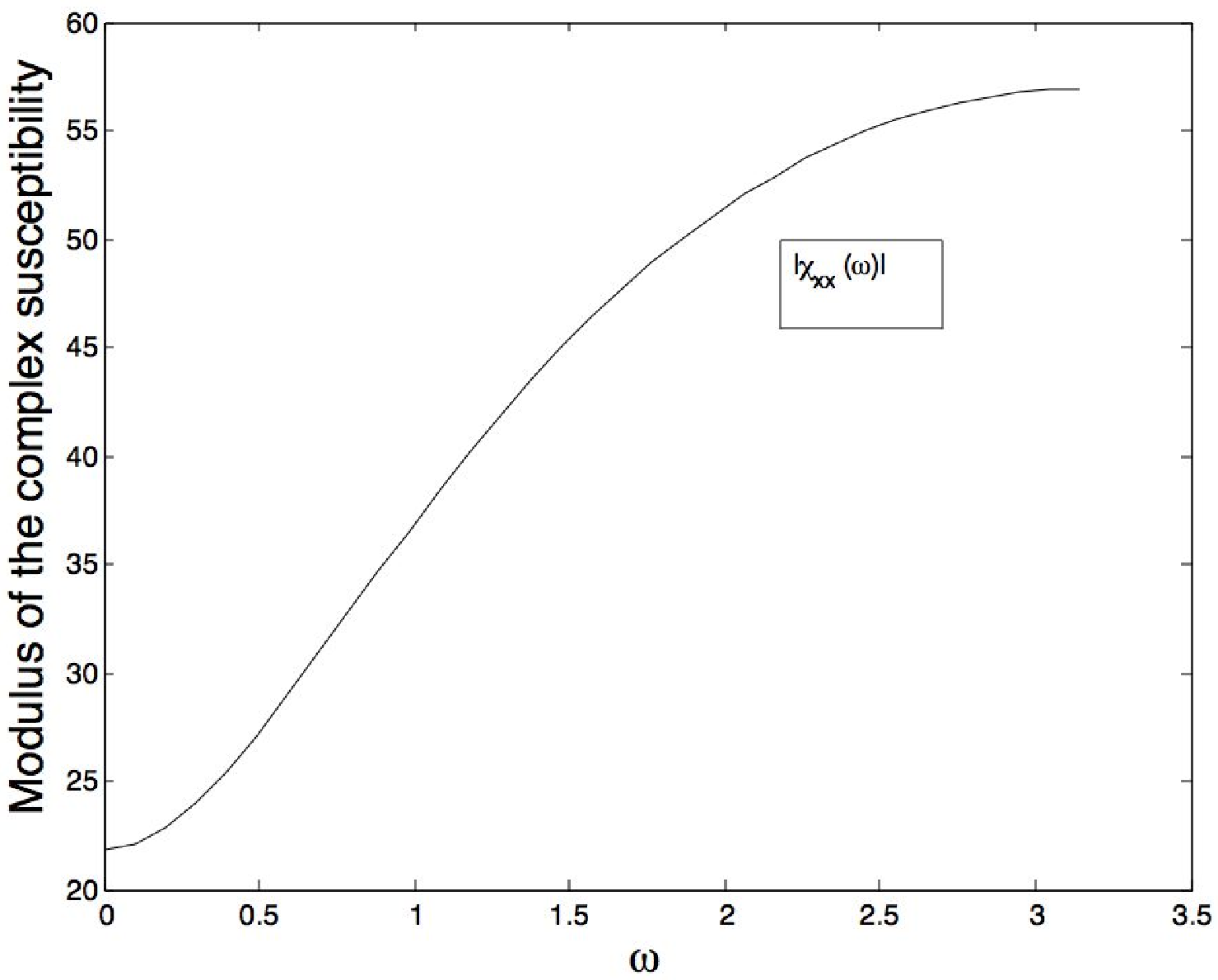}
\includegraphics[width=10cm,clip=false]{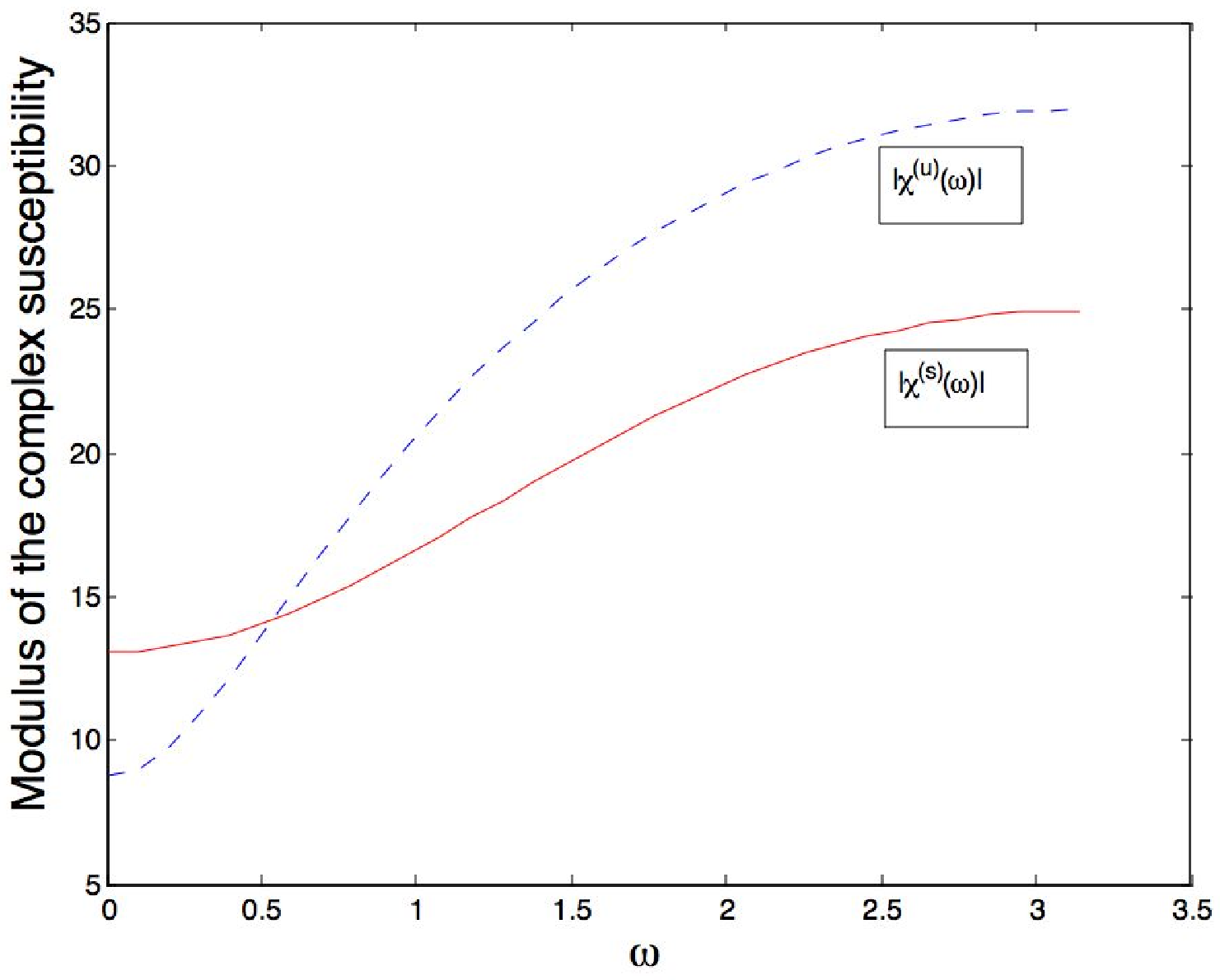}
\vspace{- 3cm}
}
\caption{\label{Sus1}  Complex susceptibility $\hat{\chi}_{xx} (\omega) $  computed from eqs.\rfq{chixx^u}-(\ref{eq:Cxx})  with parameters $\alpha = 0.8, \omega_0 = 1$. (a)  Modulus of the  complex susceptibility. (b)  Modulus of the stable (solid line)  and of the  unstable (dashed-line) susceptibility.}
\end{figure}

\begin{figure}[!ht]
\vspace{- 1cm} 
\centerline{
\includegraphics[width=10cm,clip=false]{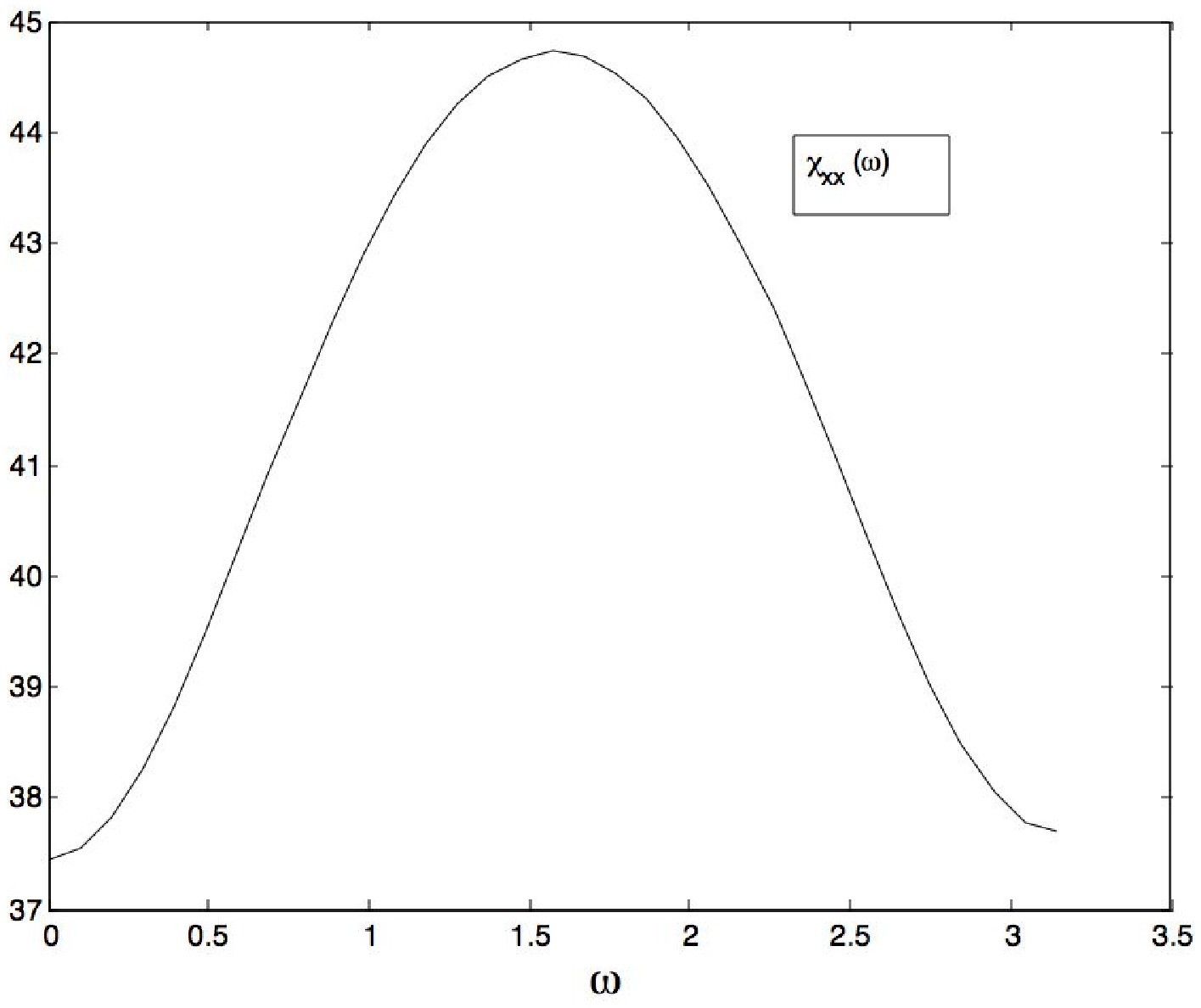}
\includegraphics[width=10cm,clip=false]{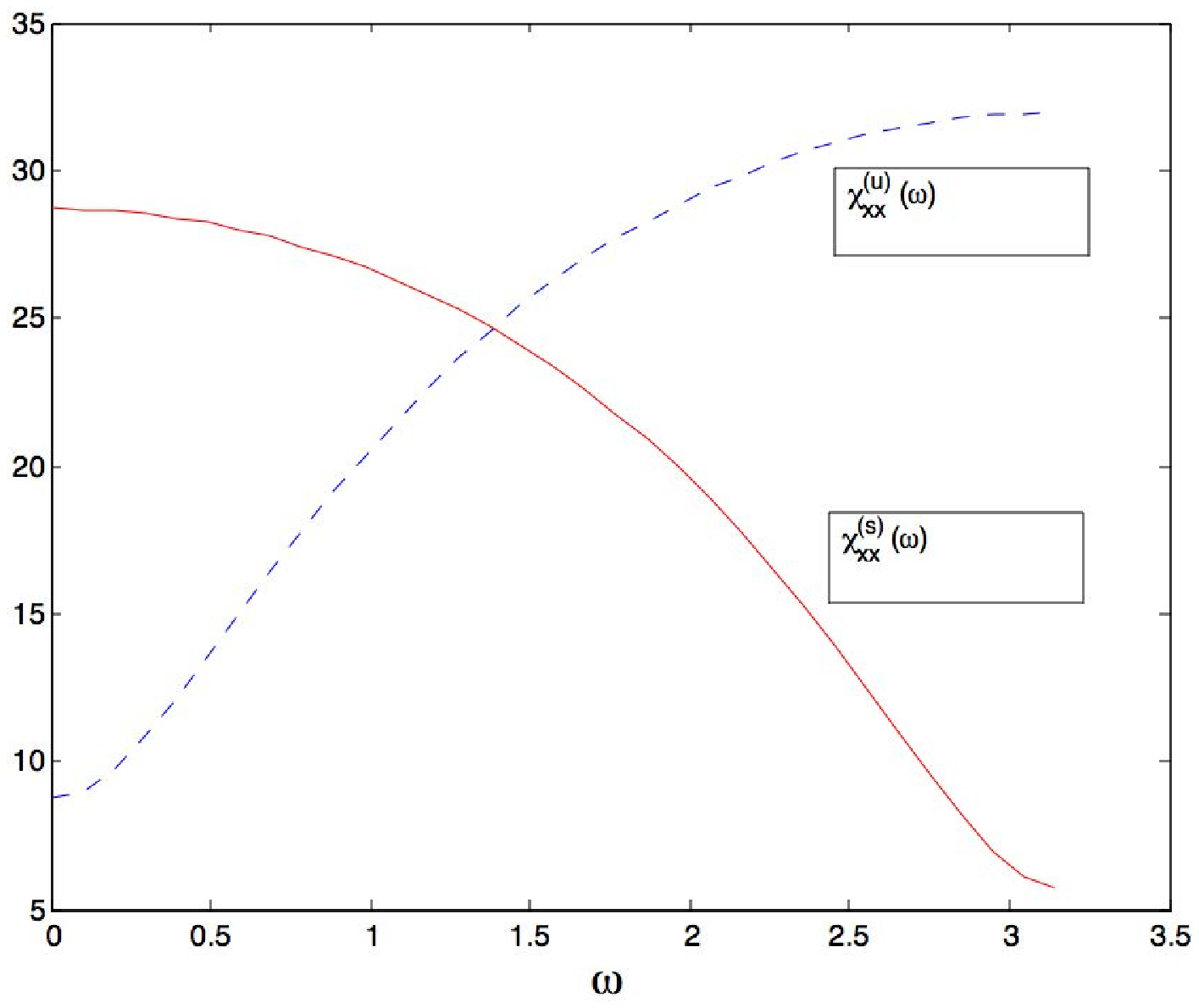}
\vspace{- 3cm}
}
\caption{\label{Sus2}  Complex susceptibility $\hat{\chi}_{xx} (\omega) $  computed from eqs.\rfq{chixx^u}-(\ref{eq:Cxx})  with parameters $\alpha = 0.8, \omega_0 = 3$. (a)  Modulus of the  complex susceptibility. (b)  Modulus of the stable (solid line)  and of the  unstable (dashed-line) susceptibility.}
\end{figure}

In this example  the poles of the complex susceptibility can thus be  controlled  by changing the parameters $\mu$ and $\omega_0$ which characterise the radial (thus here ``transverse'')  dynamics.  Figure~\ref{Sus1} shows  the modulus of the complex susceptibility $|\hat{\chi}_{xx} (\omega)|$, as well as the respective contributions of the stable and the unstable susceptibilities,  for a given choice of parameters $\alpha = 0.8, \omega_0 = 1$. 
For this choice one notices that the resonances exhibited  by these curves are all situated in $\omega = \pi$.  So in this case the information given by the Fourier transform of a correlation function, would suffice to characterise, at least qualitatively, the susceptibility of the system. In particular, the maximum response of the system to periodic forcing occurs at the same frequency that the resonance of correlation functions. 
This situation  changes by tuning parameter $\omega_0$, however.  
 Figure~\ref{Sus2}(a)  shows a case where the resonance of the complex susceptibility is no longer situated in $\omega = \pi$, but near $\pi/2$.   
 
As already discussed, this  ``anomalous'' behaviour cannot be understood in the framework of the fluctuation-response theorem, but well in the  Ruelle's response theory of dissipative chaotic systems.   On the other hand it can be remarked that  the resonance of $ | \hat{\chi}_{xx} (\omega)|  $ near $\pi/2$ does not correspond to a stable nor to an unstable pole.   In fact it results from the interference of two poles, stable and unstable, which appear as distinct  resonances, which can be seen on the graph of
 $|\hat{\chi}_{xx} ^{(s)}(\omega) |$ and of $|\hat{\chi}_{xx} ^{(s)}(\omega)|$
represented  on   
Fig.~\ref{Sus2}(b).
  
So this simple example illustrates that  the linear response of the system  depends on the behaviour or the ``stable'' response function.  In particular this  can drastically  differ  from all those of  correlation functions.  Moreover, one sees that the position of the stable poles can be tuned  by  controlling  appropriate  parameters ruling the dynamical systems. 

\subsubsection*{Amplified echo of the impulse response}

We end this Section by reporting another property of the stable response which is an application of the analytical results\rfq{chixxu}-(\ref{eq:chixys}).  The following result is also linked to the interpretation by mean function $B(t)$ [eq.\rfq{chitr1}] of  the transient  growth of response functions, as discussed in Section~\ref{trmodel}.

The response function $ \chi_{xy}^{(s)} (t)$  is a product  of an exponential function and of a correlation function.  If the latter is of amplitude of order (1) at time $t$, the result gives an amplification of $2^t$. 
 To illustrate this property consider a perturbation of the form:
\begin{equation}
X (\theta) =  \cos (2^p + 1) \theta
\label{eq:Xjparticulier3}
\end{equation}
Then is easy to work out the correlation function $< \sin \theta_t \, \sin \theta_0  X_y(\theta_0 )] >$ appearing in\rfq{chixys}.  It is found that if $p>1$, the stable response function is
\begin{equation}
\chi_{xy}^{(s)} (t)   =  -\frac{b(2^p-e^{-\mu p})}{4(2-e^{-\mu})} \, \delta_{t,p}
  \label{eq:chixysPULSE}
\end{equation}
Notice that in this case  the susceptibility  $\hat{\chi}_{xy}^{(s)} (\omega)  \propto e^{ip\omega}$ has no poles.   However, it is seen in this example that an effect of the stable direction is to produce an exponentially  amplified "echo" of the perturbation.

\section{Conclusion}

In the recent theory developed by Ruelle~\cite{Ru99}   the  standard concepts of linear response, susceptibility and resonance are revised and extended due to the dynamical  contraction of the whole  phase space onto attractors.
 More technically,  the contraction rate of phase space  leads one to replace the conservative  Liouville measure (which is the standard measure in classical statistical physics) by the notion of SRB measure.      
Among others, this change demands to go beyond   the  standard frawework  of  the  ``fluctuation-response''  theorem, which means that one cannot any longer deduce the statistical  behaviour of the system just by looking at its  correlation functions.
For example,  the powerspectra of  classical (conservative) physical systems
provide in general all the  informations about the resonances of this  system.  For the dissipative systems considered by Ruelle, this information is incomplete, however, as they may be new types of resonances --called the stable resonances because they are related to stable manifolds-- which cannot be detected in the spectra of correlation functions but in the complex susceptibility of this system.  
In previous papers we have exhibited a model of dynamical network with dissipative chaotic dynamics where  we succeeded to numerically compute  the susceptibility and show indeed the presence of new resonances in the susceptibility which are absent  in the Fourier transform of the correlation functions. 
In order to  gain  more intuitions on what bring  about these  ``stable'' resonances and their consequences, we have  considered   in   the present article simpler models which could be  treated analytically.  Moreover the analytical results could be used to test further our numerical method and to compare it to a previous method which existed in the litterature.  

We  have first  examined the 1-dimensional model $\theta_{t+1} = 2 \theta_t $ (mod $2\pi$) which presumably is the simplest chaotic system with Lyapunov exponent $\log 2$.  Despite its simplicity this model is instructive to consider,  as it is a non classical physical  systems  (having  a  purely expanding and non invertible dynamics).  The linear response of this  
system can be worked out analytically  and the result is a correlation function.   In this case it is easy to show that this response decays exponentially with a characteristic time equal to the mixing time.  Also the latter is clearly associated with a pole of the susceptibility and also identified with one  Ruelle-Policott resonance. 

Next we have studied a class of   2-dimensional models describing simple rotator models.  The angular dynamics is the same as previously but there is also a  radial dynamics which contracts the phase space  on the unit circle (fig.~\ref{fig:phaseportrait}).   The resulting system has thus a non-fractal chaotic attractor, which moreover is uniformly hyperbolic.  This elementary model could be used to illustrate several new properties associated with the dissipative  dynamics.   First it is easy to show in it   the presence of  new poles whose real parts  correspond  again  to  zero-frequency resonances but  imaginary parts are  associated with the relaxation time of  the contracting dynamics.  In fact  it is seen that the cross-response  $\hat{\chi}_{\theta r}(\omega)$   has a pole which   combines   the two characteristic  times (mixing and contracting)  in a linear combination of both. This property is demonstrated to exist in a class of more general and higher dimensional systems, as presented  in the Appendix.    In the rotator model  one could also describe the  stable resonance of  $\hat{\chi}_{\theta r}(\omega)$ by means of the  error function $B(t)$,  eq.\rfq{chitr1},   showing  how  the analytic properties of the perturbation function $X(\theta)$  regulates  the amplitude growth and decay of the response. 
Next we extended the calculations  obtained for a single  rotator to results for  a {\em network}  of coupled rotators.   This generalisation  pointed out to  the possibility to get  stable resonances wich are specific to the pair $ij$  in the network and having arbitrary non zero frequencies. Both of these  features  can be shown to be  intimately linked  to the spectral properties of the connectivity matrix of the network.  
This properties would be worth to analyse further  as they are also related to the ability of the system to propagate signals between units of the network~\cite{CS2}. 

We have also considered an extension of the rotator model in another direction which allows  one to discuss  the relation between the return time to equilibrium  of the system after an impulse perturbation and  the mixing time.   These characteristic times coincide when there is no attractor in the phase space but here we proposed a mechanism to get  a relaxation time arbitrary long in case of transverse perturbation to the attractor.  Roughly speaking it is enough  that the transverse dynamics becomes less contracting with the distance to the attractor leading possibly to the loss of hyperbolicity for critical perturbations. More precisely we have implemented such mechanism in model\rfq{trsinmodel} which presumably is beyond the scope of analytical treatment, but could be dealt with numerical simulations. These numerical results indeed showed the narrowing of stable resonances following our qualitative predicitions. 

Finally we have analysed the simple rotator model in cartesian coordinates, illustrating a rare case where the decomposition of the susceptibility $\hat{\chi}_{ij}(\omega) = \hat{\chi}_{ij}^{(s)}(\omega)+\hat{\chi}_{ij}^{(u)}(\omega)$ in respectively stable and unstable components can be performed analytically.
   This example showed in particular that in some cases a maximum in the amplitude  $|\hat{\chi}_{ij}(\omega)|$  may not  be interpretable  simply as a stable or an unstable pole,  but rather  as an interference effect of both.

The results obtained in this paper gives  more insights  in our   previous work concerning the linear response of a chaotic  network model~\cite{CS1}.  This preceding study had allowed us to numerically demonstrate the existence of the new resonances predicted by Ruelle in a model of neural network.  Thanks to a  numerical  method devised to compute the susceptibility and  to extract also some poles of it, 
these new resonances were observed with  the unexpected features to be often narrow and pair $ij$-dependent.   This last property  enabled  us also to propose a method of  signal transmission  bewteen the  nework units by means of amplitude modulation of harmonic signals locked on  the resonance frequencies~\cite{CS2}.
Now the present work permits   us to understand  some  mechanisms  able to create such resonances with arbitrary frequencies and arbitrary widths, and  the  possibility to get  pair-selective resonances. 
In future works we will examine further how the control  this pair selectivity and how  pole positionning can be achieved so that we can propose more advanced  schemes for communication between units within a chaotic network.  We anticipate  that such schemes would be quite useful  not only in the field of neural networks but also for applications in other biological networks (e.g. gene regulatory networks) or in  communication networks.  In particular we will examine also ways  of  transmitting  trains of periodic pulses, which are  prevalent  in the considered applications.   On the other hand, to demonstrate the practical relevance of these new ideas we are working on  experimental devices with dissipative chaotic dynamics  in order to produce experimental evidences of the present linear response theory~\cite{CS3}. 

From the theoretical side it would be worth to deepen the consequences  of the fractal (cantor-like) structure of generic strange attractors, a feature which was not considered in this paper.   A  generalisation of our rotator model would be to replace the contracting dynamics along the radial direction by a baker map.  
  To this purpose a good model candidate  to be investigated  in the future  could   be the nonlinear multibaker map analysed in~\cite{GFD}.   This reference considers a class of perturbed multibaker map mimicking a field driven, thermostated random walk  for which a non trivial  SRB measure can be analytically computed to first order in the field parameter. This system exhibits indeed a fractal structure in the stable direction.  Moreover in some limit,  this system becomes nonhyperbolic, in a similar way to the example numerically treated in Section 4.3. 

Finally,  many questions addressed above would be worth to be studied beyond the linear response regime, which is in the scope of  our present numerical techniques.

\appendix

\section{Some generalisation  in $n$-dimensional space} 

In this appendix we analyse further the way the  impulse  response  function $\chi_{ij} (t) $, defined by  eq.\rfq{chi_ij},  can be decomposed as the sum $\chi_{ij} (t) = \chi^{(s)}_{ij} (t) +\chi^{(u)}_{ij} (t)$ [eq.\rfq{decomp}]
 in a  higher dimensional space.  The unstable contribution $\chi^{(u)}_{ij} (t)$ in this sum has been dealt with  in details by Ruelle who introduces the concept of unstable divergence.  In what follows we recall this concept and we present an attempt of dealing with the stable part $\chi^{(u)}_{ij} (t)$.  To this purpose we introduce  hypothesis on the underlying dynamical system so that we can generalise some conclusions obtained in this paper with the simple  rotator model.

The unstable part $\chi_{ij} (t)$  has been studied   by Ruelle in~\cite{Ru99} in the case of a general observable $A$.  When $A=x_i$   the result  of this analysis  can be expressed by a  correlation function which can be written as:
\begin{eqnarray}
\chi^{(u)}_{ij} (t)  & =  &    - \int \rho_F (d\bx)  \,   F^t_i (\bx) \,  \div^u \bX_j^{u} (\bx)       \nonumber \\ 
& =  &-   <  x_i(t) ;\,  \div^u \bX_j^{u} (\bx(0))>     \label{eq:chi^u} 
\end{eqnarray}
In this equation  the differential operator $\div^u$ is called the {\em unstable divergence}, i.e. the usual
 divergence\footnote{\label{divu}   The divergence $\div_{\zeta}  \bX = \frac{1}{\zeta} \div (\zeta \bX) $  of vector field $\bX$ computed with respect to the invariant measure $\rho(d\bx) = \zeta(\bx) d\bx$ can be  defined as the Lie derivative 
$L_{\bX} ( \rho )  = ( \div_{\zeta}  \bX) \, \rho$.  } operator
 restricted to differentiating the field $ \bX_j^{u}$  only along the unstable directions and computed with respect to the invariant measure $ \rho_F (d\bx)$ (which is absolutely continous in these directions).
We propose to retrieve  the result\rfq{chi^u}   in the particular case where we can assume a change of variables
\begin{equation}
\bx = \bxi (\bu, \bs)
\label{eq:u-s}
\end{equation}
such that  $\bu$ and $\bs$ are coordinates representing  respectively the unstable and the stable manifolds of system $\bx_{t+1} =  \bF (\bx_t)$. Moreover let us suppose that coordinates  $(\bu,\bs)$  can be characterised by a {\em diagonal}  metrics:
\begin{equation}
\| d\bx \|^2 =  \sum_{k=1}^{m} g_k (\bu,\bs) du_k^2 +  \sum_{k=1}^{n} h_k (\bu,\bs) ds_k^2 \quad\quad\quad\quad (m+n = N)
\label{eq:metrics}
\end{equation}
with $g_k = \| \frac{\partial \bxi}{\partial u_k} \|^2$ and $h_k = \| \frac{\partial \bxi}{\partial s_k} \|^2$.     These hypothesis are a direct generalisation of the rotator model studied in Section~\ref{sec:rotator}. In that case the coordinates $(\bu,\bs)$  are simply the polar coordinates $(\theta,r)$.

Using these coordinates, we assume also that the invariant measure $\rho_F(d\bx)$ can be factorised as
\begin{equation}
\rho_F(d\bx) =   \rho^{(u)}(du) \,  \rho^{(s)}(ds)
\label{eq:factorise_rho}
\end{equation}
where   the measure in the stable directions, $\rho^{(s)}(ds)$,   is typically singular, and $ \rho^{(u)}$  is absolutely continuous along the unstable directions with a density $ \zeta(\bu)$.  So  the invariant measure $\rho_F(d\bx)$ will be also  written as:
\begin{equation}
\rho_F(d\bx)   =   \zeta(\bu) \, \sqrt{g (\bu,\bs)} du_1 \cdots  du_m \,  \rho^{(s)}(ds)
\label{eq:rho^u}
\end{equation}
where $\sqrt{g }$ is the determinant of the metric tensor (restricted to the unstable coordinates). 
Now we   compute the unstable contribution of the susceptibility :
\begin{equation}
\chi^{(u)}_{ij}(t)  =  \int \rho _F (d\bx)   \nabla F_i ^t  (\bx) \,\cdot\, \bX^{(u)}_j (\bx) 
\label{eq:chi_u}
\end{equation}
where $\bX^{(u)}_j $ is the projection on the unstable space of vector  $\bX_j $  (the perturbation  along 
the coordinate $x_j$).
We can assume that    $\bX^{(u)}_j$  decomposes as:
\begin{equation}
\bX^{(u)}_j (\bu,\bs) = \sum_{k=1}^{m}  X_{jk}  (\bu,\bs)   \frac{\partial \bxi}{\partial u_k}
\label{eq:Xuj}
\end{equation}
Then by  noticing that  $ \nabla F_i ^t  (\bx) \cdot \, \bX^{(u)}_j =  \sum_{k=1}^{m}   \frac{\partial F^t_i}{\partial u_k}   X_{jk} $  the unstable susceptibility can be written as:
\[
\chi^{(u)}_{ij}(t)  =  \int  \rho^{(s)}(ds) \,  \zeta \,  \sqrt{g} \, du_1 \cdots du_m 
\left(  \sum_{k=1}^{m}   \frac{\partial F^t_i}{\partial u_k}   X_{jk}   \right)
\]
Integration by part on variable $u_k$  simplifies the equation because the integrated term can be assumed to vanish. Therefore one obtains:
\begin{equation}
\chi^{(u)}_{ij}(t)  =  - \int  \rho^{(s)}(ds) du_1 \cdots du_m   F^t_i   \sum_{k=1}^{m}  \frac{\partial}{\partial u_k }  \left[  \zeta  \, \sqrt{g}   X_{jk}   \right]
\label{eq:Xuj_integrated}
\end{equation}
Finally the equation\rfq{chi^u} is retrieved by defining
\begin{equation}
\div_{\zeta}^u \bX^u =  \frac{1}{\zeta} \frac{1}{\sqrt{g}} 
 \sum_{k=1}^{m}  \frac{\partial}{\partial u_k } \left[  \zeta  \, \sqrt{g}   X_{jk}   \right]
\label{eq:unstable_div}
\end{equation}
When $\zeta=1$ the righthand side of this equation is the standard expression of the  divergence expressed  in curvilinear coordinates. The upperscript {\em u}  indicates however that the differentiation acts only with respect to the coordinates $\bu$ of the unstable manifold (see footnote \ref{divu}). 

\subsubsection*{The stable response function}
\label{sec:stable}

We focus now on the stable contribution of the decomposition\rfq{chiij} of the susceptibility. This term is thus given by:
\begin{equation}
\chi^{(s)}_{ij}(t)  =  \int \rho _F (d\bx)   \nabla F_i ^t  (\bx) \cdot \, \bX^{(s)}_j (\bx) 
\label{eq:chi_s}
\end{equation}
where $\bX^{(s)}_j $ is the projection on the stable space of the perturbation $\bX_j $.
Let us remark  that  $\bX^{(s)}_j $ can be computed  in terms of the function $\bxi$ as:
\begin{equation}
\bX^{(s)}_j  = X_j \sum_{k=1}^{m}    \frac{1}{h_k} 
  \frac{\partial \xi_j}{\partial s_k}   \frac{\partial \bxi}{\partial s_k} 
\label{eq:Xsj}
\end{equation}
%
So  the integral for  $\chi^{(s)}_{ij}$ becomes sightly more explicit:
\begin{equation}
\chi^{(s)}_{ij}(t)  =  \int \rho^{(s)}(ds) \rho^{(u)}(du)     \sum_{k=1}^n \left(  \frac{1}{h_k}  \frac{\partial F_i ^t} {\partial s_k}  X_j      \frac{\partial \xi_j } {\partial s_k}  \right) 
\label{eq:chi_s_bis}
\end{equation}
The   reason why we cannot  transform  this expression into a correlation function, as was done for the unstable part, is that here we are not allowed to  integrate by part on variables $s_k$.  Indeed the invariant measure is typically singular in $\bs$ (there is no density function in $\rho^{(s)}$).  At this stage one cannot proceed further in the development of $ \chi_{ij}^{(s)} (t) $ unless additional hypothesis are made on the dynamics.

In what follows, simplifying  hypothesis  will be added  on the dynamics in order to factorise the integral\rfq{chi_s_bis} into separate contributions of respectively the stable and the unstable coordinates.  Although these assumptions  are compelling, they are  satisfied in the case of our simple  rotator model, thus generalising  the properties obtained in this framework. 

So let us consider  four factorisation   hypothesis as follows.  The first one is that the dynamics $\bx_{t+1} = \bF (\bx_t)$  decouples in terms of coordinates $(\bu,\bs)$ and  becomes
\begin{eqnarray}
\bu_{t+1} & = &  \bU(\bu_t)  \nonumber \\
\bs_{t+1} & = &  \bS(\bs_t)  \label{eq:us}
\end{eqnarray}
with some functions $\bU$ and $\bS$.   The second assumption  is that the change of variables from $(\bu,\bs)$ to $\bx$ can be written as the product
\begin{equation}
x_j = \xi_j (\bu,\bs) =  \varphi_j(\bu)\, \eta_j( \bs).
\label{eq:xiBIS}
\end{equation}
By combining these two assumptions   the $t$-th iterates of dynamics can  factorise   as $F_i^t(\bx) =  \varphi_i(\bU^t(\bu))\, \eta_i( \bS^t(\bs))$. 
The third hypothesis is that the metrics functions  $h_k$ defined in eq.\rfq{metrics} can be also factorised:
\[
h_k (\bu,\bs) = \alpha_k (\bs) \, \beta_k(\bu) 
\]
Then, assuming finally the following choice for the perturbing function $X_j(\bx) = \Gamma(\bs) \psi (\bu)$, the stable susceptibility can be factorised as:
\begin{eqnarray}
\chi^{(s)}_{ij}(t)  & =  &  \sum_{k=1}^n \left(   \int \rho^{(s)}(ds)    \frac{1}{\alpha_k}  \frac{\partial \eta_i}{\partial s_k} (\bS^t(\bs) )   \frac{\partial \eta_j}{\partial s_k} (\bs)   \Gamma (\bs)  \right) 
 \left(   \int  \rho^{(u)}(du)    \frac{1}{\beta_k}  \varphi_i(\bU^t(\bu)) \varphi_j ( \bu)  \psi(\bu) \right) 
  \nonumber \\
& = &  \sum_{k=1}^n Z_{ijk}(t) \, C_{ijk} (t)  \label{eq:chi^sZC}
\end{eqnarray}
In this last expression of $\chi^{(s)}_{ij}(t) $,  the $C_{ijk}(t)$ are correlation functions of the unstable dynamics, namely
\begin{equation}
C_{ijk}(t) =  <\varphi_i (\bu_t) ;\, \varphi_j (\bu_0) \frac{\psi (\bu_0)}{\beta_k(\bu_0)}>
\label{eq:Ck_t}
\end{equation}
whereas the $Z_{ijk}(t)$ are not correlation functions because it cannot be expressed as\rfq{correlation}.

In order to simplify the notation, now  we consider the case where there is only one stable direction ($n=1$). Then eq.\rfq{chi^sZC} reduces to:
\begin{equation}
 \chi^{(s)}_{ij}(t)  =  Z_{ij}(t) \, C_{ij}(t)
 \label{eq:chi^simple}
 \end{equation}
For hyperbolic mixing systems  the correlation function $C_{ij}(t)$ can be decomposed as a sum:
\[
C_{ij}(t) = \sum_{k=1}^{\infty}  c_{ijk} \, e^{- (\mu_k+i \omega_k) t}  + c.c.
\]
whose Fourier transform is readily deduced as:
\[
\hat{C}_{ij}(\omega) = \sum_{k=1}^{\infty}  \frac{c_{ijk}}{1- e^{i (\omega-\omega_k) - \mu_k}} + c.c.
\]
This function possesses complex poles for $\omega = \pm \omega_k -i \mu_k  $. These are the Ruelle-Policott resonances already mentioned and they do not depend on the choice of the pair $ij$.
For hyperbolic mixing systems one gets $0 < \mu_1 \leq \mu_2 \leq \cdots$ and $\mu_1^{-1}$ is the mixing time (cf. Section~\ref{sec:mixing} ).  Thus the Fourier transform of\rfq{chi^simple} becomes  the convolution product:
\begin{equation}
\hat{ \chi}^{(s)}_{ij}(\omega)  =  \frac{1}{2\pi} \int \! \!  \frac{c_{ijk} \, \hat{Z}_{ij}(\omega^{\prime} )}{1- e^{i (\omega -\omega^{\prime} -\omega_k) - \mu_k}}  d\omega^{\prime}
 \label{eq:chi^simple1}
\end{equation}
where $\hat{Z}_{ij}(\omega)$ is the Fourier transform of  $Z_{ij}(t)$.   The latter is not a correlation function but  it  can be assumed to  exponentially  converge  towards $0$  since it is related to the stable dynamics.
So,  due to the singularities of $\hat{Z}_{ij}(\omega)$  the complex susceptibility $\hat{ \chi}^{(s)}_{ij}(\omega) $ can acquire new poles distinct from the Ruelle-Policott resonances, and the latter can depend on the pair $ij$ as mentionned in Section~\ref{sec:network}.    For example let us  consider the simplest case for $Z_{ij}(t)$  where it is  a function of the form: 
\begin{equation}
Z_{ij}(t) =  a_{ij} \, e^{- (q_{ij}+i p_{ij}) t}  + c.c.     \quad \quad \quad \quad (t \geq 0).
\label{eq:Zij}
\end{equation}
Then the complex susceptibility takes the expression:
\begin{equation}
\hat{ \chi} ^{(s)}_{ij}(\omega)  =   \sum_{k=1}^{\infty} \left(  \frac{a_{ij} \, c_{ijk}}{1- e^{i (\omega-\omega_k-p_{ij}) - \mu_k-q_{ij}}} + \frac{\overline{a}_{ij} \, c_{ijk}}{1- e^{i (\omega-\omega_k+p_{ij})  - \mu_k-q_{ij}}} \right) + c.c.
\label{eq:chi^simple2}
\end{equation}
Therefore, in this  simple case where $\hat{Z}_{ij}(\omega)$ can be characterised by only two poles $q_{ij}\pm ip_{ij}$, it is seen that the stable resonances are obtained by splitting  the Ruelle-Policott resonances in any  combinations of the form
$\pm \omega_k \pm p_{ij} -i (\mu_k + q_{ij})$.   Notice that  the real parts of these new poles are bounded from below by $\mu_1$.  So in this framework the time of return to equilibrium  is  the ``mixing'' time.

Extending  the example of eq.\rfq{Zij},  new  resonances can be created  for each poles of  $\hat{Z}_{ij}$ and for each terms (i.e. each stable directions) in the more general expression\rfq{chi^sZC} of $\chi^{(s)}_{ij}(t) $.  Therefore we conclude that the existence of  these poles enables the susceptibility to possess resonances which possibly are specific to the pair $ij$. This means that a periodic forcing of one  degree of freedom $j$ of the system can result in principle in a  resonance in  another degree of freedom $i$  whose frequency  is specific to it.  This is precisely this  possibility of $ij$-dependent resonances which is exploited  in~\cite{CS2} in order to perform selective transmission of signals  in a chaotic networks of interconnected units.



\begin{thebibliography}{00}
\bibitem{topol}  The general definition of the SRB measure
 $\rho_F$ is $\lim_{T\rightarrow \infty} \frac{1}{T} \sum_{t=1}^{T} \bF^{t*} m$ 
but it is equivalent to $\rho_F = \lim_{t \rightarrow \infty}  F^{*t} \; dx$ if $F$ is topologically mixing.
\end{thebibliography}
 \end{document}